\documentclass[preprint,nohyper]{JHEP3} %
\usepackage{epsfig,multicol}
\usepackage{latexsym}
\usepackage{amssymb,amsmath}
\usepackage{amsfonts}
\usepackage{multirow,slashbox}
\usepackage{graphics,subfigure,graphicx}

\newcommand{\dis}[1]{\begin{equation}\begin{split}#1\end{split}\end{equation}}
\newcommand{\disn}[1]{\begin{equation*}\begin{split}#1\end{split}\end{equation*}}

\newcommand{\be}{\begin{equation}}
\newcommand{\ee}{\end{equation}}

\newcommand{\p}{\partial}

\newcommand{\Tb}{{\bar{T}}}
\newcommand{\Cb}{{\bar{C}}}
\newcommand{\sech}{\,\textrm{sech}}

\newcommand{\eprint}[1]{\href{http://arxiv.org/abs/#1}{#1}}

\newcommand{\adsurl}[1]{\href{#1}{ADS}}
\providecommand{\url}[1]{\href{#1}{#1}}

\title{Moduli Evolution in the Presence of Matter Fields and Flux Compactification}
\author{Carsten van de Bruck \\Department of Applied Mathematics \\ University of Sheffield \\Sheffield, S3 7RH. UK \\\email{C.vandeBruck@sheffield.ac.uk}}
\author{Ki-Young Choi \\Department of Physics and Astronomy \\ University of Sheffield \\Sheffield, S3 7RH. UK \\\email{k.choi@sheffield.ac.uk}}
\author{Lisa M.H. Hall \\Department of Applied Mathematics \\ University of Sheffield \\Sheffield, S3 7RH. UK \\\email{lisa.hall@sheffield.ac.uk}}

\abstract{We provide a detailed analysis of the dynamics of moduli fields in the KKLT scenario coupled to a Polonyi field, which plays the role of a hidden matter sector field.
It was previously shown that such matter fields can uplift AdS vacua to Minkowski or de~Sitter vacua.
Additionally, we take a background fluid into account (which can be either matter or radiation), which aids moduli stabilisation.
Our analysis shows that the presence of the matter field further aids stabilisation, due to a new scaling regime.
We study the system both analytically and numerically.
}

\preprint{}
\begin{document}

\section{Introduction}
Theories beyond the standard model, such as string theory, predict the existence of 
massless (or nearly massless) scalar fields. These so-called moduli fields may contain, for example, 
the information about the dynamics of extra spatial dimensions, or, as it is the 
case of the string theory dilaton, they determine the coupling strength of gauge fields. 
In addition, these fields usually interact, with gravitational strength, to other particles, 
thereby mediating a new (``fifth'') force. Since so far we have neither observed any 
significant time-dependence of gauge couplings nor detected new forces, the moduli fields have 
to be stabilised~\cite{Dine:1985he}.  This should happen at some stage during the cosmological evolution, which is one of 
the problems addressed in string cosmology. It was emphasized in \cite{Steinhardt} that 
stabilising the string theory dilaton in a cosmological framework is a rather difficult, because of the 
the small barrier height separating a local Minkowski (or de Sitter) vacuum and the steepness 
of the potential. The situation can be improved if background fluids (either radiation or matter) 
or temperature dependent corrections are taken into account (for work in this direction see e.g.~\cite{Kaloper,Barreiro1,Barreiro2,Huey:2000jx}).

In order to stabilise the moduli fields, a mechanism is required which gives them a 
mass. Additionally, the resulting vacuum energy should be very small and non-negative and 
the vacuum itself quasi-stable (i.e.~with a life-time much longer than the age of the universe). 
Recently, compactification mechanisms with fluxes on internal manifolds have been 
extensively studied (see \cite{Giddings:2001yu}, a review can be found in \cite{fluxreview}); 
a well-investigated scenario is the KKLT proposal \cite{KKLT}. 
Within this latter setup, moduli fields are stabilised and the resulting cosmological 
constant can be fine-tuned to be either zero or very small. To obtain 
a realistic scenario, a crucial ingredient in these mechanisms is a ``lifting'' 
procedure, which raised a supersymmetric AdS-vacuum to become a Minkowski or even 
a (quasi) de-Sitter vacuum. In the original KKLT proposal, an anti-D3 brane was added, 
explicitly breaking supersymmetry. In an alternative proposal, D-terms were considered to 
up-lift the AdS-vacuum, which requires the existence of charged matter fields \cite{Burgess,Nilles,Carlos}. 

Recently it was pointed out that interactions of the moduli fields with a hidden matter 
sector can also result in an effective uplifting to Minkowski or de Sitter vacua. The idea 
of this procedure is to take matter fields into account (such fields will be 
present in any realistic theory). The resulting interaction terms in the scalar potential 
lead to spontaneous supersymmetry breaking, with local minima which have a small and 
positive vacuum energy density. This procedure is known as F-term uplifting (for 
recent work, see see e.g.~\cite{Silverstein,Lebedev,GomezReino,Hebecker,okklt,Dudas,Brax,abe} 
and references therein). An appealing feature is that this procedure can result in a 
small gravitino mass, while maintaining a high potential barrier, resulting in long-lived 
de-Sitter vacua \cite{Lebedev,okklt,Dudas,Lebedev2}.

In this paper, we investigate the cosmological dynamics of the moduli fields and matter fields 
in the the context of F-term uplifing. As a toy model, we will couple the KKLT model with the Polonyi 
model \cite{Polonyi}, following \cite{Lebedev,abe,Lebedev2,abe2}. This model is simple enough to study 
the dynamical consequences of the interactions between the moduli fields and matter fields, but we believe that our results can be generalised to more complex models. In particular 
we are interested in whether the interaction will facilitate or impede the stabilisation of the moduli 
fields in the cosmological context. The paper is organized as follows: In Section~\ref{secBack} we formalise the set-up and write down the essential equations.  In Section~\ref{secPotential}, we discuss the properties of the potential.  We review past results, in the context of our potential, in Section~\ref{sec_review_single} and generalise to multi-field scenarios in Section~\ref{secReal} (two real fields).  The full dynamics for two complex fields is presented in Section~\ref{secComplex}.  We briefly consider the effects of a radiation background fluid in Section~\ref{secRad}.  We conclude in Section~\ref{secConc}.
We give further useful formulae in the Appendices and derive the new scaling regime.

\section{Background Equations}
\label{secBack}
We begin by stating the four-dimensional $N=1$ SUGRA action, which is of the form
\dis{
{\cal S}=-\int \sqrt{-g}\left(\frac{1}{2\kappa_P^2}R +K_{i\bar{j}}\partial_\mu
\Phi^i\partial^\mu\bar{\Phi}^{\bar{j}} +V \right)d^4x,
}
where $K_{i\bar{j}}=\frac{\partial^2 K}{\partial \Phi^i\partial \bar{\Phi}^{\bar{j}}}$ is 
the K\"ahler metric, $\Phi^i$  are complex chiral superfields and $V(\Phi)$ is 
the scalar potential. $\kappa^2_P$ is the 4-dimensional Newton constant,
\dis{
\kappa^2_P=8\pi G_N=1.
}
The effective scalar potential is given, with given K\"ahler metric $K$ and superpotential $W$, as
\dis{
V=e^K\left(K^{i\bar{j}}D_i W \overline{D_jW}-3 W\overline{W} \right),
}
where $K^{i\bar{j}}$ is the inverse K\"ahler metric and 
$D_i W=\partial_i W+ \frac{\partial K}{\partial \Phi^i} W$.

The following equations of motion for the real and imaginary parts of superfields are~\cite{Barreiro:2005ua}
\dis{
&\ddot{\varphi}^i_R+3H\dot{\varphi}^i_R +\Gamma^i_{jk}(\dot{\varphi}^j_R\dot{\varphi}^k_R-\dot{\varphi}
^j_I\dot{\varphi}^k_I)+\frac12 K^{i\bar{j}}
\partial_{j_R}V=0,\\
&\ddot{\varphi}^i_I+3H\dot{\varphi}^i_I +\Gamma^i_{jk}(\dot{\varphi}^j_I\dot{\varphi}^k_R+\dot{\varphi}
^j_R\dot{\varphi}^k_I)+\frac12 K^{i\bar{j}}
\partial_{j_I}V=0,
}
where $\varphi^i_R(\varphi^i_I)$ refers to the real (imaginary) part of the scalar 
fields and $\partial_{j_R}$ ($\partial_{j_I}$) are used to denote 
the derivative of the potential with respect to the real (imaginary) 
parts of the fields, respectively.
The connections on the K\"ahler manifold are given by
\dis{
\Gamma^n_{ij}=K^{n\bar{l}}\frac{\partial K_{j\bar{l}}}{\partial \Phi^i}.
}

There are two different sectors in the model considered here; one involves a 
modulus field $T$, whereas the other sector contains a matter field $C$. Both fields 
will be taken as complex fields. The corresponding K\"ahler potential~\cite{Lebedev2},
which arises in type IIB and heterotic string theory, is 
\dis{
K=& -3\ln(T+\bar{T})+|C|^2,\\
W=& \mathcal{W}(T)+ \mathfrak{W}(C).
}
Following~\cite{Lebedev2,abe2}, for example, we consider the combination of the KKLT~\cite{KKLT} 
and the Polonyi model, for which the superpotential is given by
\dis{
\mathcal{W}(T)&=W_0+Ae^{-aT},\\
\mathfrak{W}(C)&=c+\mu^2 C,
}
so that the resulting scalar potential is given by 

\dis{
V=& \frac{e^{C\Cb}}{3(T+\Tb)^3}
\left(\left|-Aa e^{-aT}(T+\Tb)-3(W_0+Ae^{-aT}+c+\mu^2C)\right|^2\right.\\
&\qquad\qquad +\Big\lvert\mu^2\Cb+(W_0+Ae^{-aT}+c+\mu^2C)\Big\rvert^2
-3\Big\lvert W_0+Ae^{-aT}+c+\mu^2C\Big\rvert^2\Bigg).
}
The differential equations describing the dynamics of the field $T$ are given by
\dis{
\ddot{T_r} + 3H\dot{T_r}-\frac{1}{T_r}(\dot{T_r}^2-\dot{T_i}^2)+ \frac{2T_r^2}{3}\p_{T_r}V=&0,\\
\ddot{T_i}+3H\dot{T_i}-\frac{2}{T_r}\dot{T_r}\dot{T_i}+\frac{2T_r^2}{3}\p_{T_i}V=&0,
}
where we write $T=T_r+iT_i$. Furthermore, we consider a background fluid with density 
$\rho_b$ and with equation of state $\gamma-1 \equiv p_b/\rho_b$.  Energy conservation dictates that
\begin{eqnarray}
\dot{\rho_b}+3H\gamma \rho_b=&0.
\end{eqnarray}
We consider $C$ to be a complex field, $C=C_r+iC_i$, so that the equations of motion 
can be written as 
\dis{
\ddot{C_r}+3H\dot{C_r}+\frac12\p_{C_r} V=&0,\\
\ddot{C_i}+3H\dot{C_i}+\frac12\p_{C_i} V=&0.
}
The Friedmann equation reads 
\dis{
3H^2=\frac{3}{4T_r^2}(\dot{T_r^2}+\dot{T_i}^2)+
(\dot{C_r}^2+\dot{C_i}^2)+V+\rho_b.
}
Since $T_r$ is not a canonical field, it is convenient to define a new field $\phi$

\dis{
\phi\equiv \sqrt{\frac32}\ln T_r.
}
Additionally, the equations for the fields are easier to solve if we define a new set of variables 
as follows
\dis{
x\equiv \frac{\dot{\phi}}{\sqrt6 H},\qquad y\equiv \frac{\sqrt{V}}{\sqrt3 H},\qquad z\equiv\frac12e^{-\sqrt
{\frac23}\phi}\frac{\dot{T_i}}{H},
}
and
\dis{
p \equiv \frac{\dot{C_r}}{\sqrt3 H},\qquad q\equiv \frac{\dot{C_i}}{\sqrt3 H}.
}
With these variables, the Friedmann equation becomes
\dis{
1=x^2+y^2+z^2+p^2+q^2+\Omega_b,
}
where $\Omega_b=\rho_b/3H^2$.
The evolution of Hubble parameter $H$ is
\dis{
H'=-\frac32H\left[2x^2+2z^2+2p^2+2q^2+\gamma(1-x^2-y^2-z^2-p^2-q^2) \right],
}
where the prime denotes differentiation with repect to $e$-fold number $N$, defined by 
\dis{
N\equiv \int H dt.
}
The equations of motion for the fields are now given by 
\begin{eqnarray}
x'&=&-3x+\lambda\sqrt{\frac32}y^2-2z^2+\frac32x\left[2x^2+2z^2+2p^2+2q^2+\gamma(1-x^2-y^2-z^2-
p^2-q^2) \right],\nonumber \\
y'&=&-\lambda\sqrt{\frac32}xy-\eta\sqrt{\frac32}yz-\delta\frac{\sqrt{3}}{2}yp-\theta\frac{\sqrt{3}}{2}yq+
\frac32y\left[ 2x^2 + 2z^2 + 2p^2 + 2q^2 \right. \nonumber \\ 
&+& \left. \gamma(1-x^2-y^2-z^2-p^2-q^2) \right],  \\
z'&=&-3z+\eta\sqrt{\frac32}y^2+2xz+\frac32z\left[2x^2+2z^2+2p^2+2q^2+\gamma(1-x^2-y^2-z^2-p^2-q^2) 
\right], \nonumber \label{EOMxyz}
\end{eqnarray}
and
\dis{
p'&=-3p +\delta\frac{\sqrt3}{2}y^2 +\frac32p\left[2x^2+2z^2+2p^2+2q^2+\gamma(1-x^2-y^2-z^2-p^2-q^2) 
\right],\\
q'&=-3q +\theta\frac{\sqrt3}{2}y^2 +\frac32q\left[2x^2+2z^2+2p^2+2q^2+\gamma(1-x^2-y^2-z^2-p^2-q^2) 
\right].\label{EOMpq}
}
In these equations, we have defined $\lambda$ and $\eta$ to be 
\dis{
\lambda\equiv -\frac{1}{V}\frac{\p V}{\p \phi}, \qquad \eta \equiv -\sqrt{\frac23} e^{\sqrt{\frac23}\phi}\frac{1}
{V}\frac{\p V}{\p T_i},
\label{eq:lambda}
}
\dis{
\delta\equiv -\frac{1}{V}\frac{\p V}{\p C_r}, \qquad \theta\equiv -\frac{1}{V}\frac{\p V}{\p C_i}.
\label{eq:delta}
}
For the remainder of this paper, we solve the above equations, in order to investigate the stabilisation of the moduli fields, $T_r$ and $T_i$, in the presence of the matter fields, $C_r$ and $C_i$, and the background fluid, $\rho_b$.

\section{Properties of the Potential}
\label{secPotential}
As a specific example in this paper, following \cite{Lebedev2}, we take the potential parameters to be
\dis{
A=1,\qquad a=12,\qquad \mu=10^{-8},\qquad c=\mu^2\left(2-\sqrt{3}\right).
}
Additionally, $W_0$ is assumed to be real and is fine-tuned such that the potential is zero (or the 
cosmological constant $\Lambda$) at the local minimum.
For the values above, the local minimum is found at the co-ordinates
\dis{
T_r^{\rm (lm)}\simeq3.3474\,~(\phi^{\rm (lm)}\simeq1.4797)  ,\quad T_i^{\rm (lm)}=0,\quad  C_r^{\rm (lm)}\simeq-0.7131,\quad C_i^{\rm (lm)}=0.
}

\FIGURE[t!]{
  \begin{tabular}{c c}
    \includegraphics[width=0.5\textwidth]{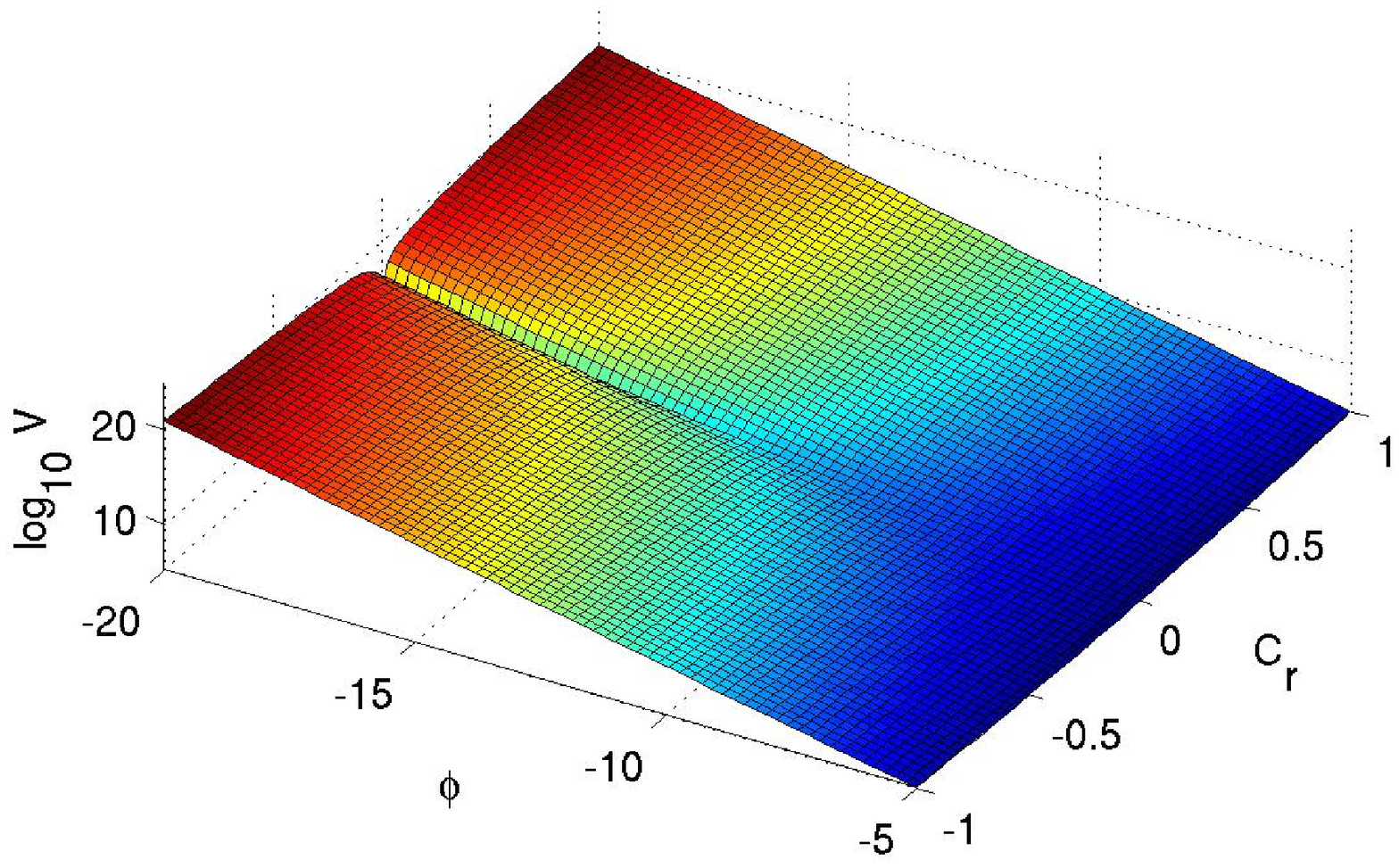}
&
    \includegraphics[width=0.5\textwidth]{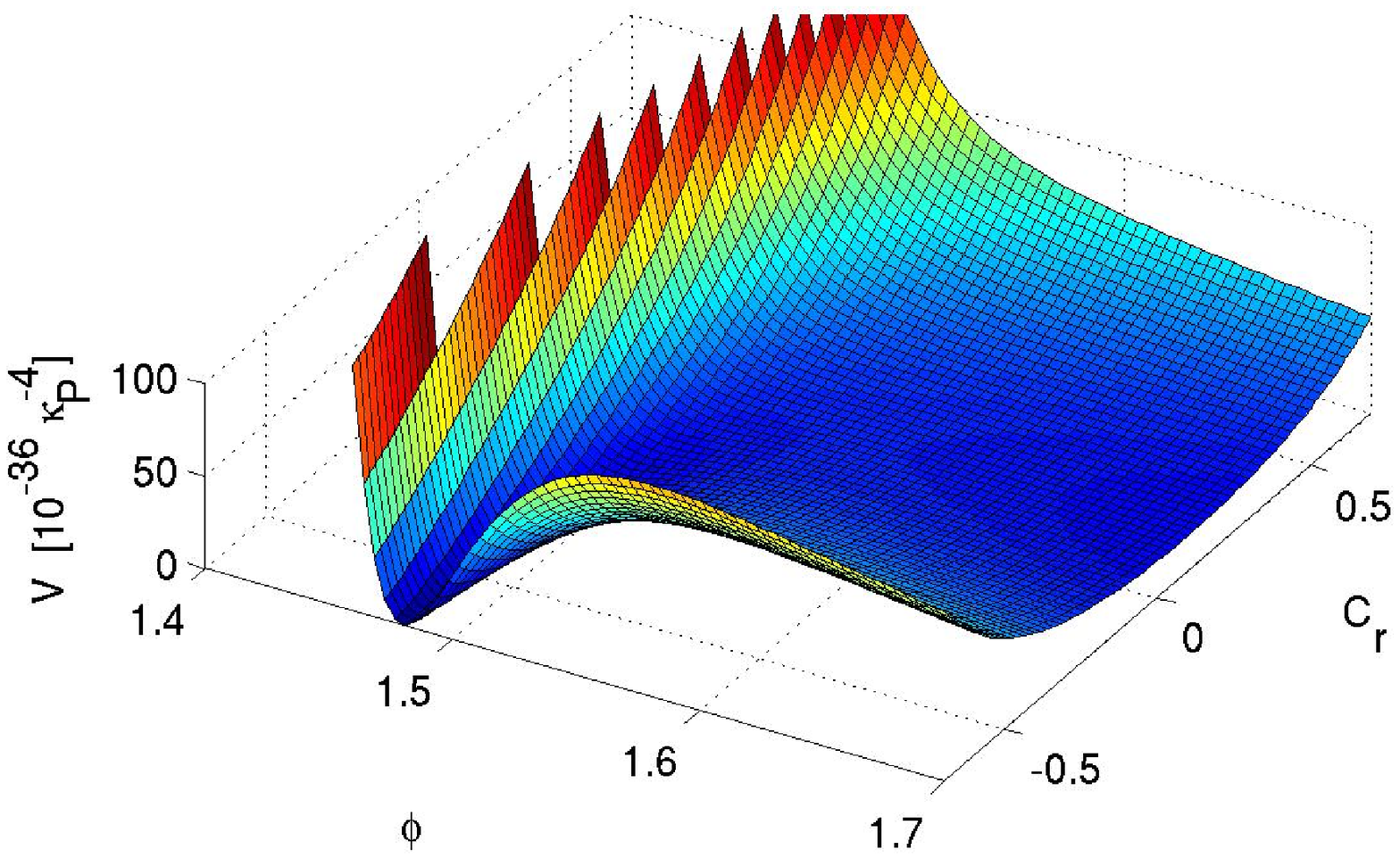}
   \end{tabular}
\caption{Scalar potential of the KKLT + Polonyi model
in units of $10^{-36}\kappa_p^{-4}$.
 We plot the potential in the $T_r$-$C_r$ plane (left hand side) and $T_r$-$T_i$ plane (right hand side),
with the other fields set at their local minimum values shown in the text. }
\label{fig-ex1}
}

\begin{figure}[t!]
    \subfigure[$\lambda$]{\scalebox{0.37}{\includegraphics{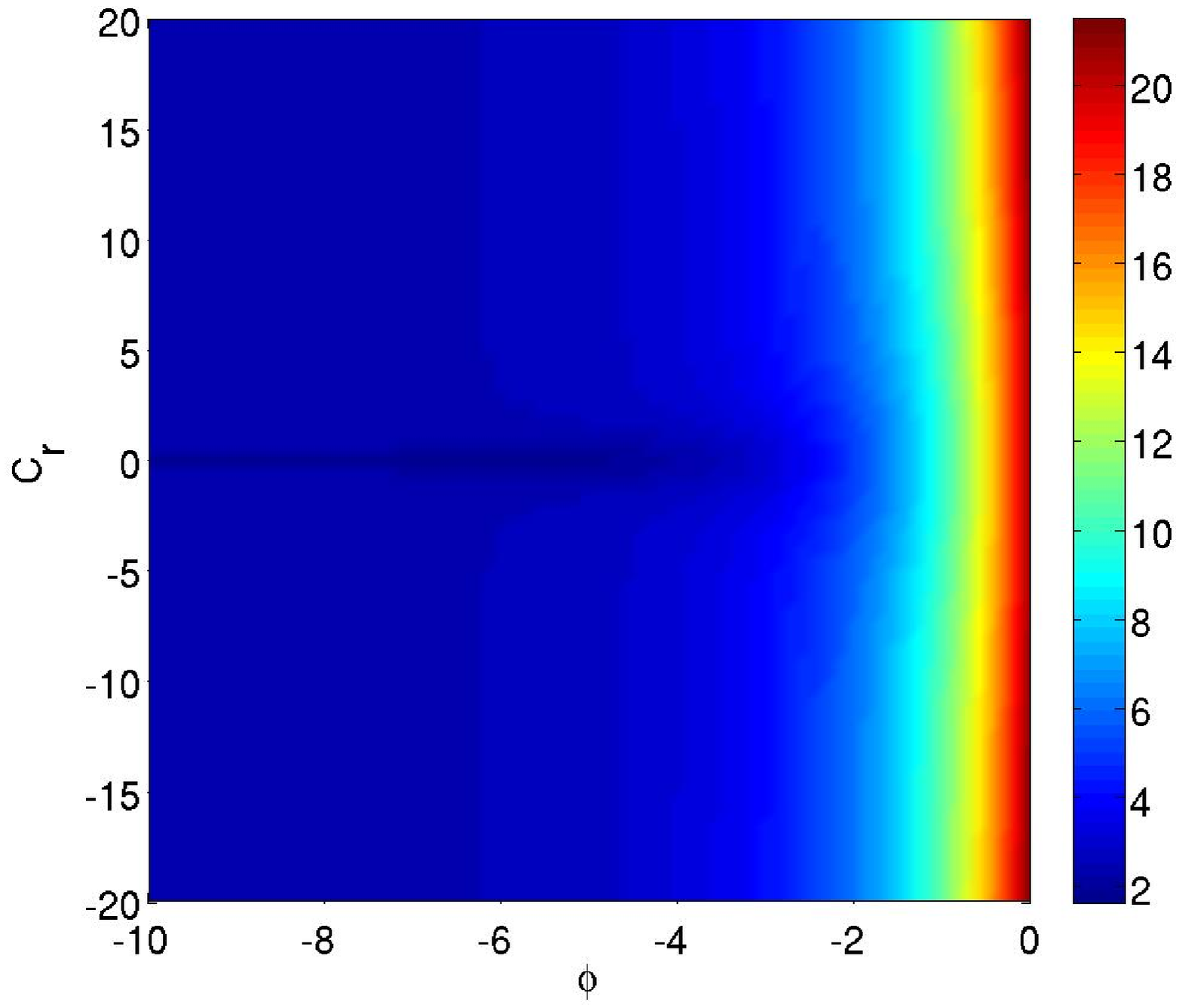}}\label{fig:lambda}}
    \subfigure[$\delta$]{\scalebox{0.37}{\includegraphics{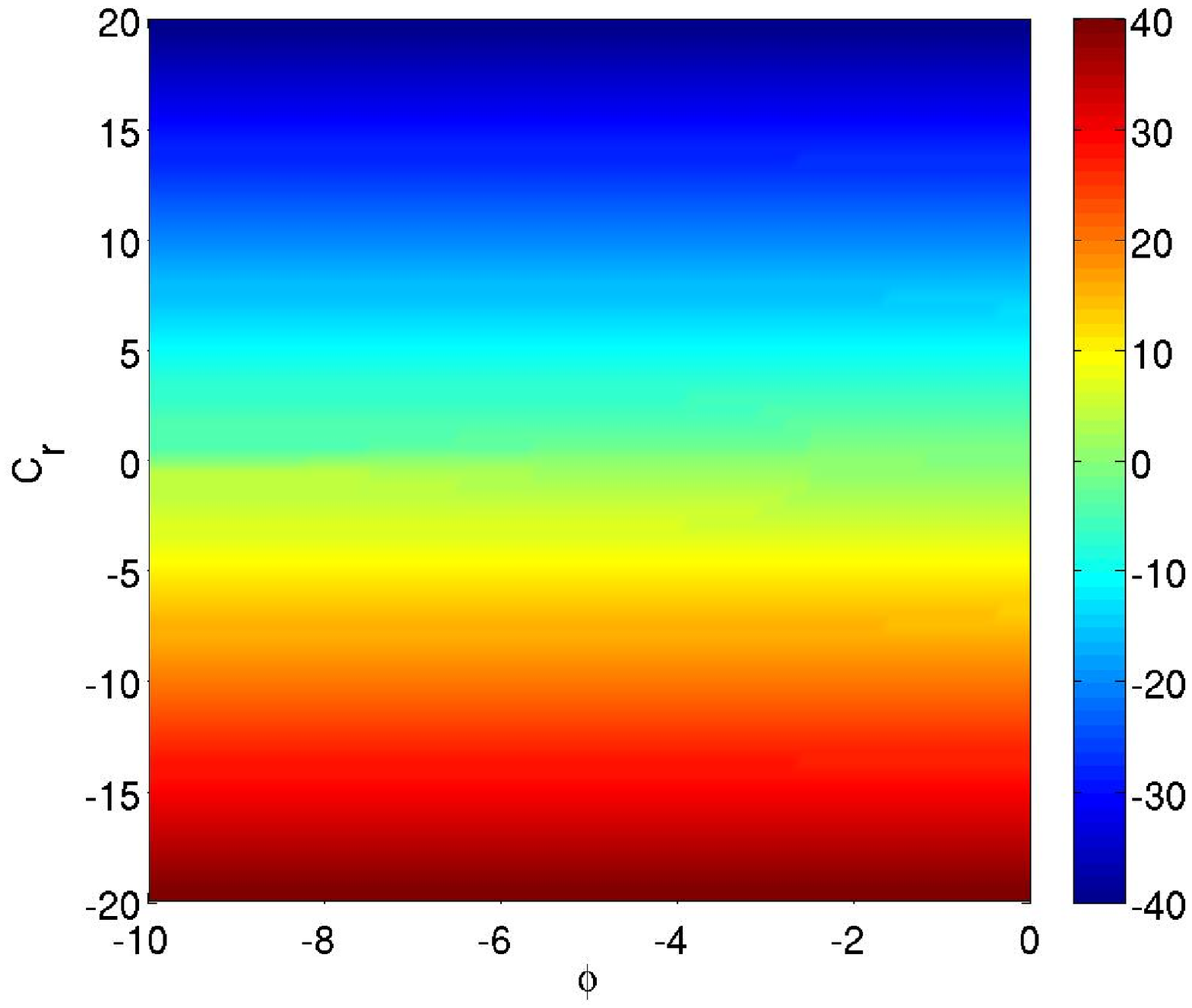}}\label{fig:delta}}
\caption{The contours of $\lambda$ and $\delta$ (defined in Eqns.~(\ref{eq:lambda}-\ref{eq:delta})) over the range of $\phi$ and $C_r$ considered in the text, for two real 
fields ($\sigma_i=C_i=0$).} 
\label{lambda+delta}
\end{figure}

\FIGURE[t!]{
  \begin{tabular}{c c}
    \includegraphics[width=0.5\textwidth]{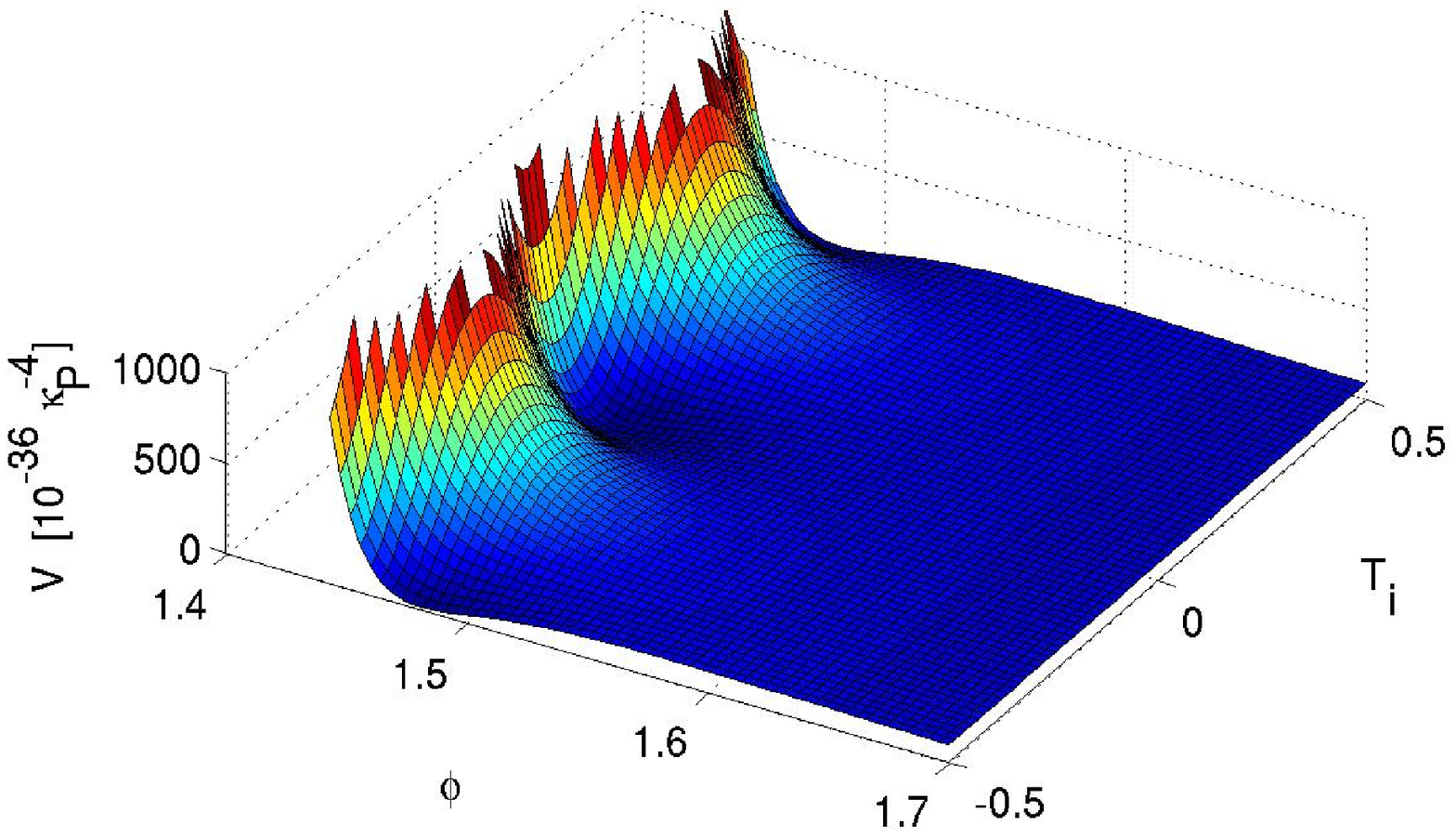}
&
    \includegraphics[width=0.5\textwidth]{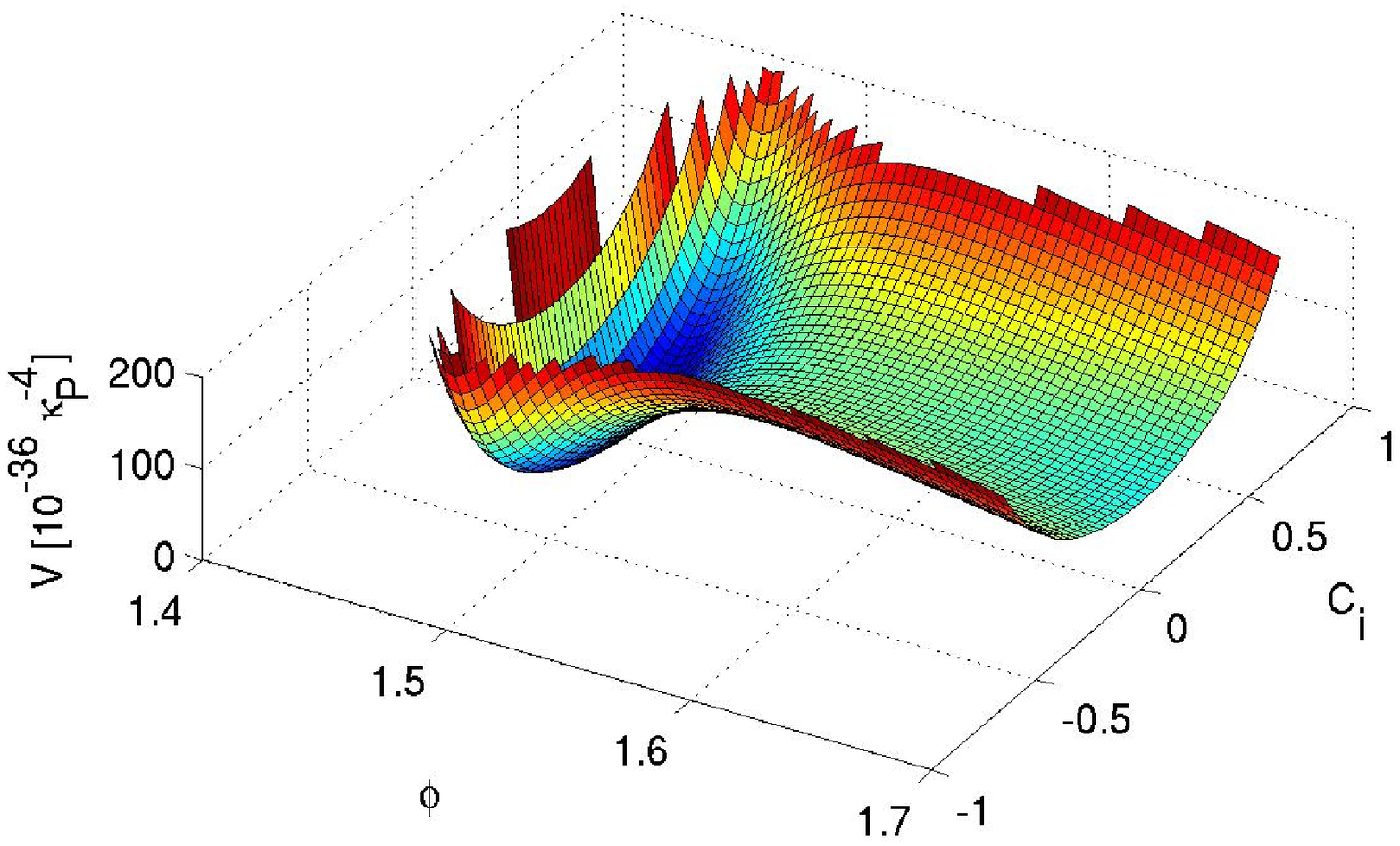}
    \end{tabular}
\caption{Scalar potential of the KKLT + Polonyi model
in units of $10^{-36}\kappa_p^{-4}$.
 We plot the potential in the $\phi$-$T_i$ plane (left hand side) and $\phi$-$C_i$ plane (right hand side),
with the other fields set at their local minimum values shown in the text. } 
\label{fig-ex2}
}
In Figs.~\ref{fig-ex1} and~\ref{fig-ex2} 
we show the shape of this potential with the parameters given above.
Since there are four independent fields, we fix two fields at their local minimum values and plot the 
remaining two fields.
For convenience, we plot the canonical field, $\phi$, instead of $T_r$.
In Fig.~\ref{fig-ex1} the local minimum of
this model is seen, where $\phi$ is stabilised at $\phi\approx 1.47$. 
As $\phi\rightarrow\infty$, the potential vanishes. 
Along the $C_r$ direction, the potential increases as $C_r$ departs from the minimum. 
As can be seen, the minimum of $C_r$ depends directly on the value of $\phi$.  
Around the local minimum, $C_r\sim -0.71$.  
For $\phi\lesssim -5$, the potential can be approximated by
\dis{
V\sim \frac{A^2C_r^2e^{C_r^2}}{8e^{\sqrt{6}\phi}}
}
with $T_i=C_i=0$, where 
the minimum of $C_r$ is $\sim 0$.  Note that the $C_r$ direction is much steeper than along the $\phi$ 
direction.
On the left hand side of Fig.~\ref{fig-ex2}, we show the sinusoidal shape of $T_i$, with periodicity of $2\pi/a$, with one minimum at zero. 
On the right hand side, the minimum of $C_i$ is seen to exist at zero and shows symmetry
about $C_i\rightarrow -C_i$ because we assume $T_i=0$ in the figure.
The values of $\lambda$ and $\delta$ (defined in Eqns.~(\ref{eq:lambda}-\ref{eq:delta})) over useful ranges are plotted in Fig.~\ref{lambda+delta}.

It should be noted that, as the $T_i$ and $C_i$ evolve away from their minima, the shape of the potential
changes from that shown in the figures.  This will become important when we consider the evolution of the imaginary parts of the fields, $C$ and $T$.

\section{Review of Single Field Dynamics}
\label{sec_review_single}

We begin by reviewing known results \cite{Barreiro1,Brustein:2004jp} and consider only real fields; 
we put the imaginary fields 
in the minimum point, $T_i=C_i=0$, so that these fields do not evolve.
This will enable isolation of the effects of each field (real and imaginary parts).  Some of the results obtained will prove vital later.

In this simplifed scenario, the potential becomes
\dis{
V=&\frac{e^{C_r^2-2aT_r}}{24T_r^3}\Big[A^2\{3C_r^2+4aT_r(3+aT_r)\}+3e^{2aT_r}(cC_r+\mu^2+\mu^2 
C_r^2+\omega C_r)^2\\
&+6Ae^{aT_r}\{c(C_r^2+2aT_r)+\mu^2 C_r^3 
+C_r(\mu^2+2a\mu^2 T_r) +\omega C_r^2+2a\omega T_r\} \Big].
\label{V_real}
}
Further, to compare to single field dynamics, for negatively large $\phi$, we set $C_r$ in its local 
minimum ($C_r=0$) initially, such that it does not evolve until the final stages of the evolution. 
Even with only $\phi$ evolving, due to the background fluid and the potential, the dynamics are non 
trivial.  
\begin{figure}[t]   
  \begin{center}
    \subfigure[]{\scalebox{0.37}{\includegraphics{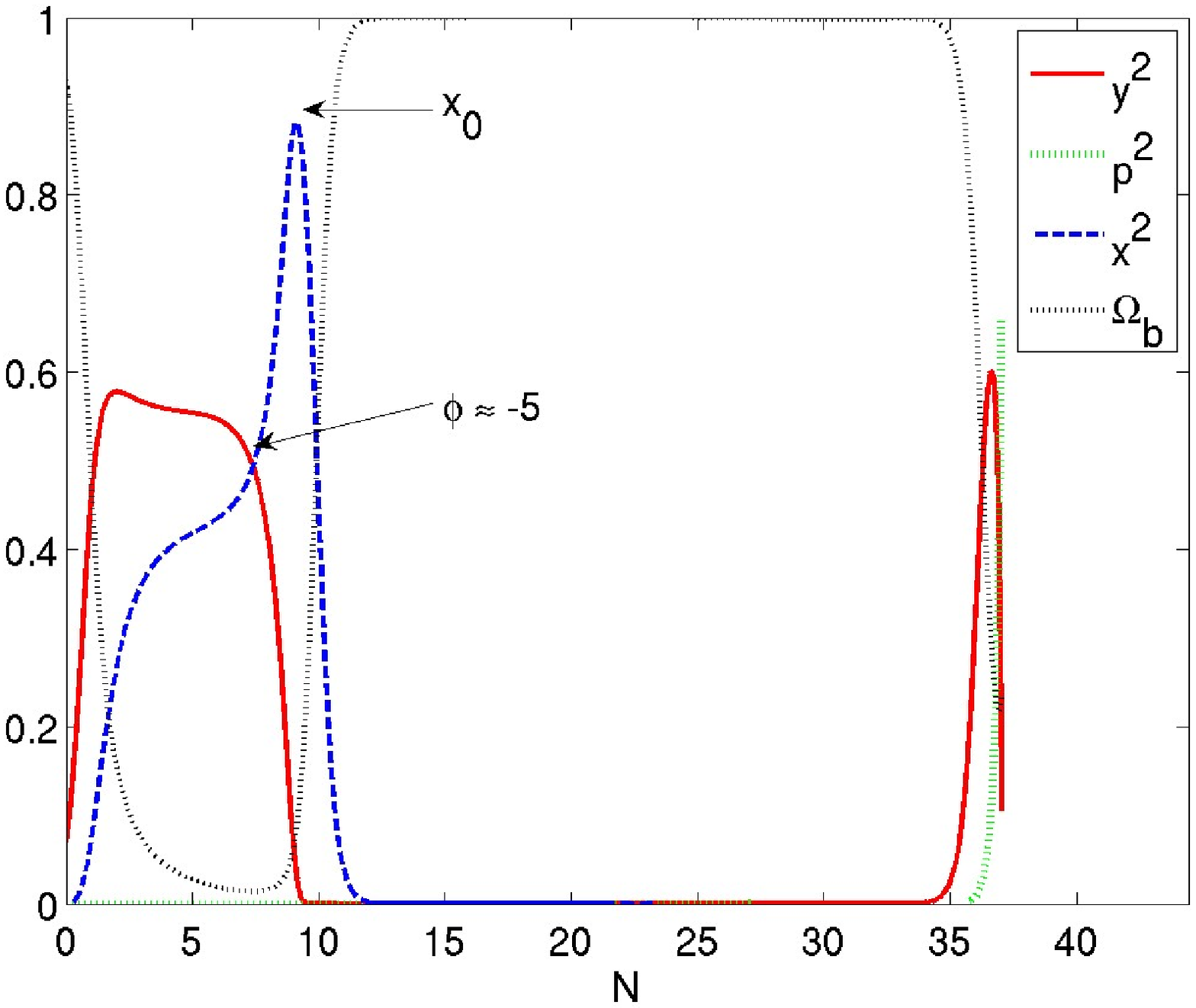}}\label{fig-single_traj_a}}
    \subfigure[]{\scalebox{0.37}{\includegraphics{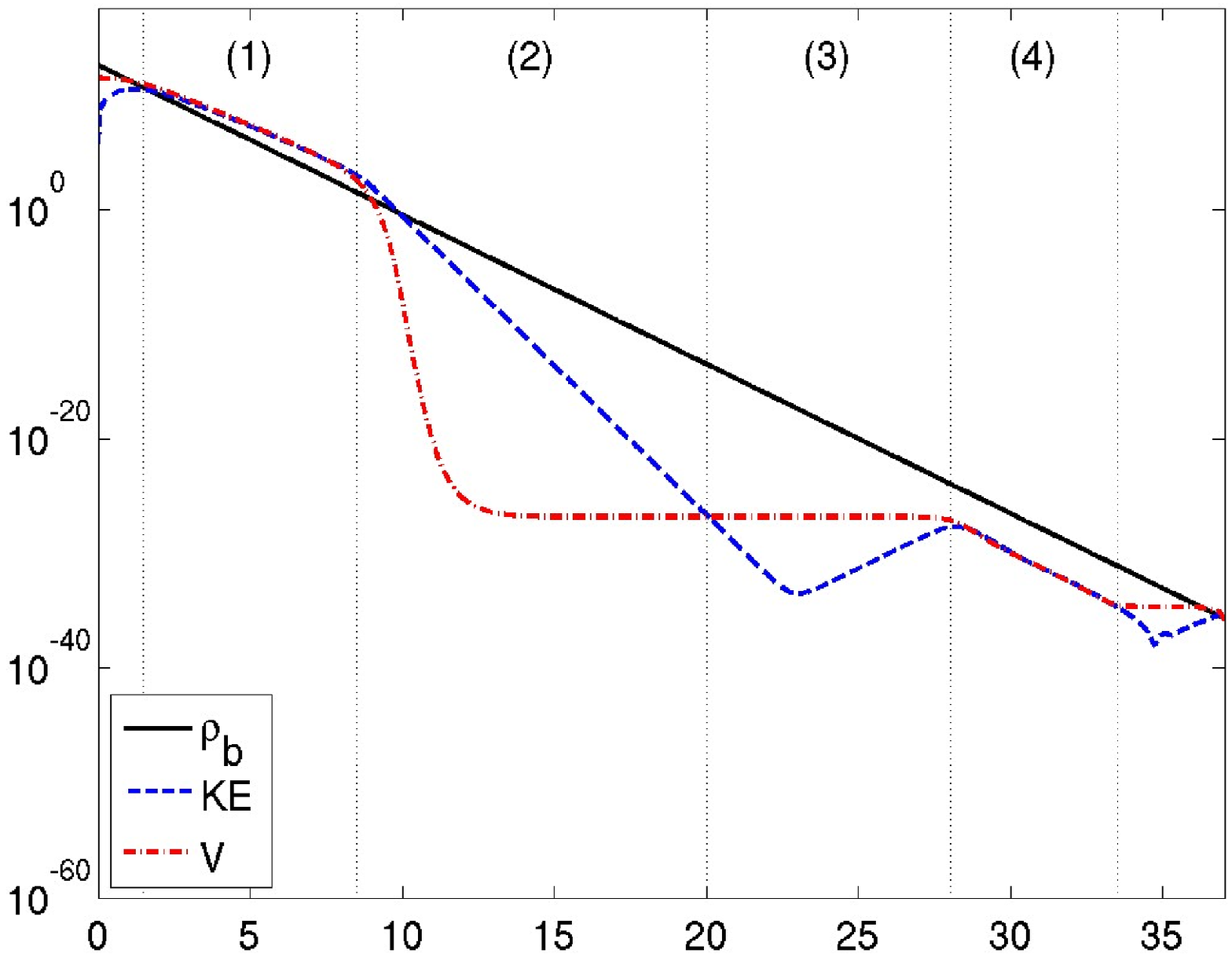}}\label{fig-single_traj_b}}
    \caption{Single Field Dynamics. Initial conditions are taken to be $\Omega_{b0}=0.93$, $C_{r0}=0$ and $\phi_0=-15$.  The background fluid is matter, with $\gamma=1$. Note that $\dot
{C_r}$ ($p^2$) is zero until the last few e-folds.}
\label{fig-single_traj}
  \end{center}
\end{figure}

In this section, we will closely follow \cite{Barreiro:2005ua} and will refer to an example trajectory shown 
in Fig.~\ref{fig-single_traj}.  
We assume that $\phi$ starts with zero kinetic energy from a 
point $\phi_{f2}$ (this terminology will become useful later) and $\Omega_{b}$ is close to unity initially ($\Omega_{b0}=0.93$).  
As the field starts to evolve, as long as $\lambda$ is small enough, the potential energy quickly
dominates the dynamics, seen in Region 1 of Fig.~\ref{fig-single_traj_b}.  
As the potential becomes more steep ($\lambda$ increases sharply around $\phi\approx -5$ for our 
potential), the evolution becomes kinetically dominated, called "kination" shown in Region 2.
Once the background fluid dominates, the field is effectively frozen at a point $\phi_{f3}$ (in Region 3), and 
the field only starts rolling again once this fluid has sufficiently diluted.  

As the field recommences to roll, one of two things can happen; either the scalar field dominates again 
(Region 1) or a scaling regime is found, in which the potential and kinetic energies track the background 
fluid (Region 4), which occurs for our trajectory. The choice of regime is specified by a scaling condition:
\begin{eqnarray}
3\gamma<\lambda^2,
\end{eqnarray}
where we restate that $\gamma=1+p_b/\rho_b$.
For our choice of potential, this condition is always satisfied when $\phi\gtrsim -5$ with a matter background fluid.
The scaling regime ends when this condition is subsequently violated, which occurs if the potential flattens, for example near the minimum.  
If $\phi_{f3}<\phi^{\rm (lm)}$, the end of scaling occurs near the minimum and the field stabilises.  If $\phi_{f3}>\phi^{\rm (lm)}$ but smaller than the local maximum value, the field will still stabilise.  Conversely, if $\phi_{f3}$ is larger than the local maximum value, the field will roll to infinity.
In our case, the fields are stabilised.

In the following, we state the analytical solutions for the single field scenario.
In 
particular, two initial branches can be distinguished; $\Omega_{b0}\ll1$ and $\Omega_{b0}\lesssim 1$
(the latter being described heuristically above).  It is convenient to consider the new variables, $x$, $y$ etc. (defined earlier) for analysis of the evolution.

\subsection{$\Omega_{b0}\ll 1$}
When $\Omega_{b0}\ll1$, following \cite{Ng:2001hs}, the initial evolution is given by
\begin{eqnarray}
x^{\prime}=\lambda\sqrt{\frac32}y^2, \qquad y^{\prime}=-\lambda\sqrt{\frac32}xy
\end{eqnarray}
where we assume $\lambda^{\prime}\approx0$, for which the solution is
\begin{eqnarray}
x&=&\sqrt{(1-\Omega_{b0})}\tanh\left(\lambda\sqrt{\frac32}N\right), 
\label{singlefielddomx} \\
y&=&\sqrt{(1-\Omega_{b0})}\sech\left(\lambda\sqrt{\frac32}N\right).
\label{singlefielddomy}
\end{eqnarray}
This solution is valid until $\lambda^{\prime}$ is no longer insignificant ($\phi\approx -5$), at which point, 
the kinetic energy starts to dominate, $x^2\simeq y^2\simeq 0.5$ and kination starts.  
Due to the non-constancy of $\lambda$, no approximation can be found for this period.  
However, during this stage, due to the insignificance of $\Omega_b$, $x^2$ increases to almost unity, as the kinetic 
energy fully dominates the evolution.  Due to the expansion of the universe, the kinetic energy is diluted 
and slowly the background fluid becomes significant ($\Omega_b\rightarrow1$, $x^2\rightarrow0$), at 
which point the field becomes frozen at $\phi_{f3}$.  

When the potential is insignificant, the system during kination is greatly simplified and can be described 
by~\cite{Barreiro1}:
\begin{eqnarray}
x^{\prime} = -3x +\frac32 x\left[2x^2+\gamma\left(1-x^2\right)\right].
\end{eqnarray} 
If $x$ starts at a value $x_0$, the solution is
\begin{eqnarray}
x=\left(1+\frac{1-x_0^2}{x_0^2}e^{3(2-\gamma)N}\right)^{-\frac12}.
\end{eqnarray}
During this decay, the field moves a distance, $\Delta\phi$, given by
\begin{eqnarray}
\Delta\phi=\frac{\sqrt{6}}{3(2-\gamma)}\ln\left(\frac{1+x_0}{1-x_0}\right).
\label{delphi}
\end{eqnarray} 
after which the field freezes at $\phi_{f3}$.  (Note that $x_0$ is never exactly unity since $\Omega_b\ne0
$).

\subsection{$\Omega_{b0}\lesssim 1$}
When $\Omega_{b0}\lesssim 1$, the the kinetic energy of $\phi$ never fully 
dominates, in that $x_0^2<1$, before the background fluid takes over.  As seen in Fig.~\ref{fig-single_traj_a}, the background fluid is never totally negligible, which complicates the evolution.
However, kination starts when $\lambda$ starts to increase ($\phi\approx -5$, $x^2\simeq 
y^2\simeq 0.5$).
Once $y^2$ is negligible, $\phi$ moves a further distance given by Eqn (\ref{delphi}) before 
freezing at $\phi_{f3}$.
Although this regime cannot be solved analytically, one point can be made: in this case, the field freezes earlier (due to less kinetic energy overall, $x_0<1
$), so that $\phi_{f3}$ is smaller than that when $\Omega_b(0)\ll1$.

Numerically, we find that, for $\gamma=1$ and $\Omega_{b0}=0.93$ (an example that will prove useful later), stabilisation only occurs when $\phi_{f2}\gtrsim-17$ and $\phi_{f3}\lesssim 1$.  We define these points as $\phi_{min}$ and $\phi_{max}$.

\section{Two Real Fields with a Matter Background Fluid}
\label{secReal}
This section extends the previous single field scenario, by considering the evolution of an additional scalar field.  Specifically, we consider the evolution of $C_r$, in which direction the potential is much steeper than the $\phi$-field previously considered.
We will show that this leads to a new mechanism for stabilisation, due to the 
presence of a new scaling regime.

\begin{figure}[t!]   
  \begin{center}
   \subfigure[]{
\scalebox{0.37}{\includegraphics{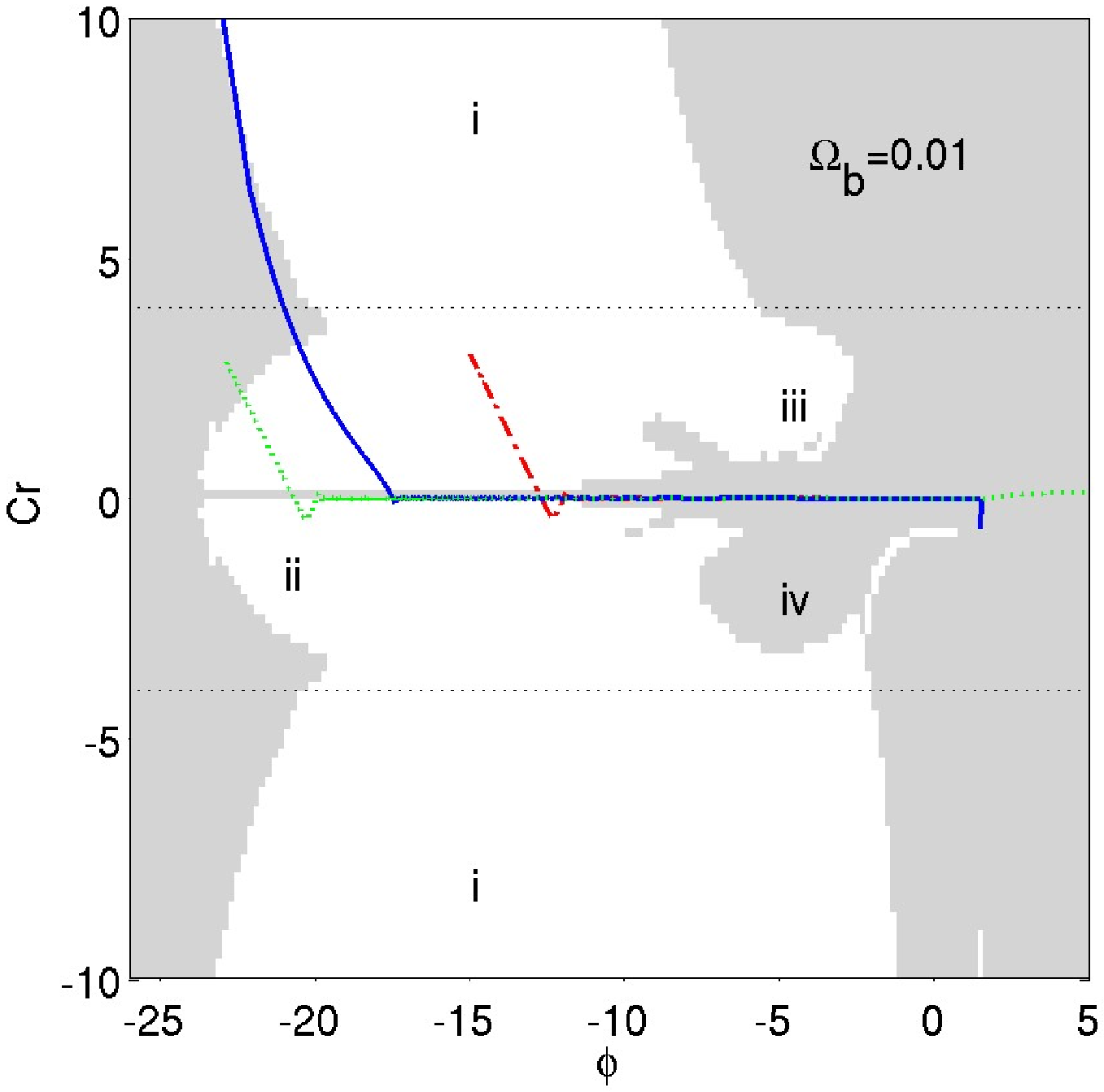}}\label{fig-real_trajs_c}}
   \subfigure[]{
\scalebox{0.37}{\includegraphics{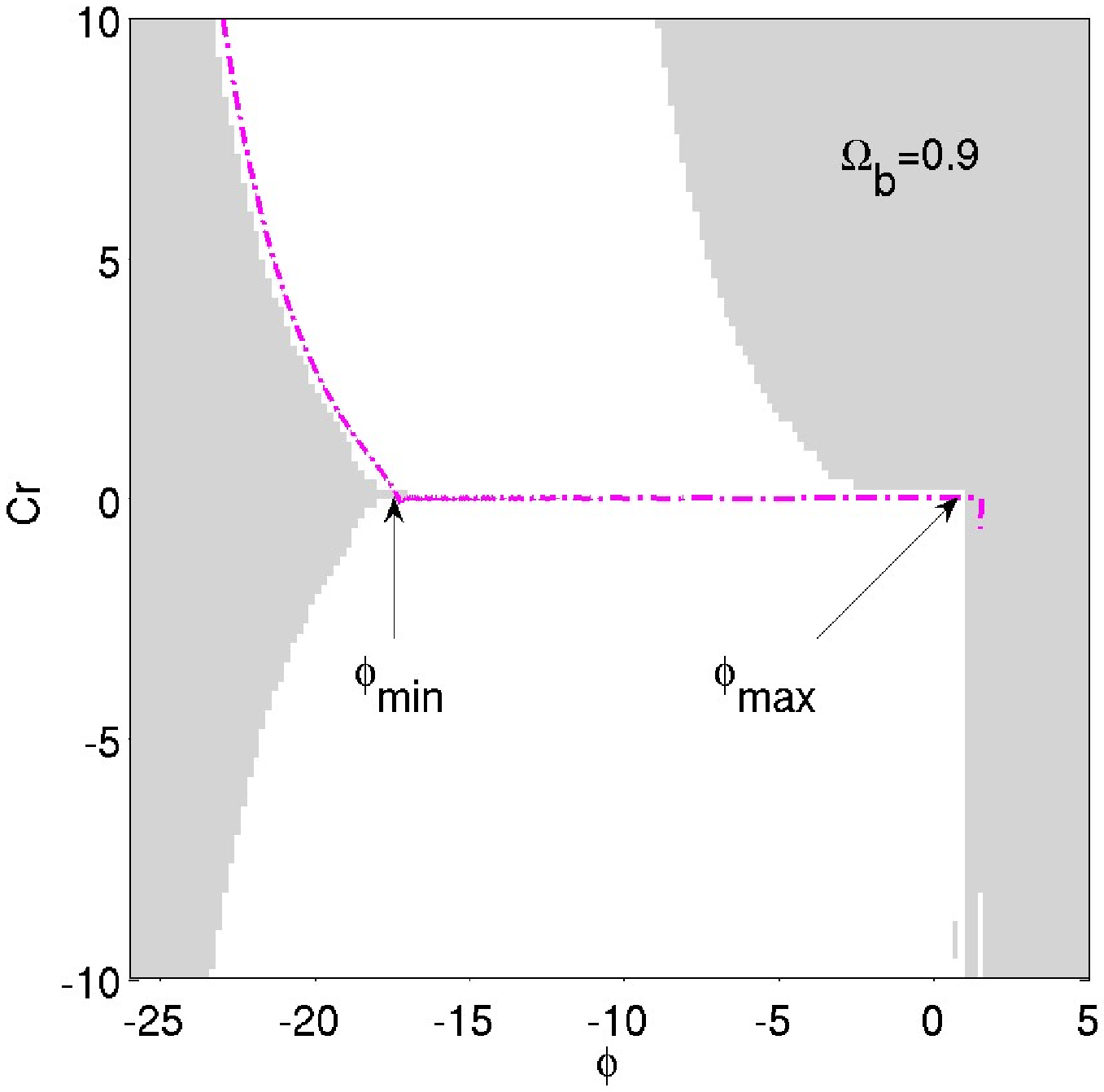}}\label{fig-real_trajs_d}}
\subfigure[]{
\scalebox{0.37}{\includegraphics{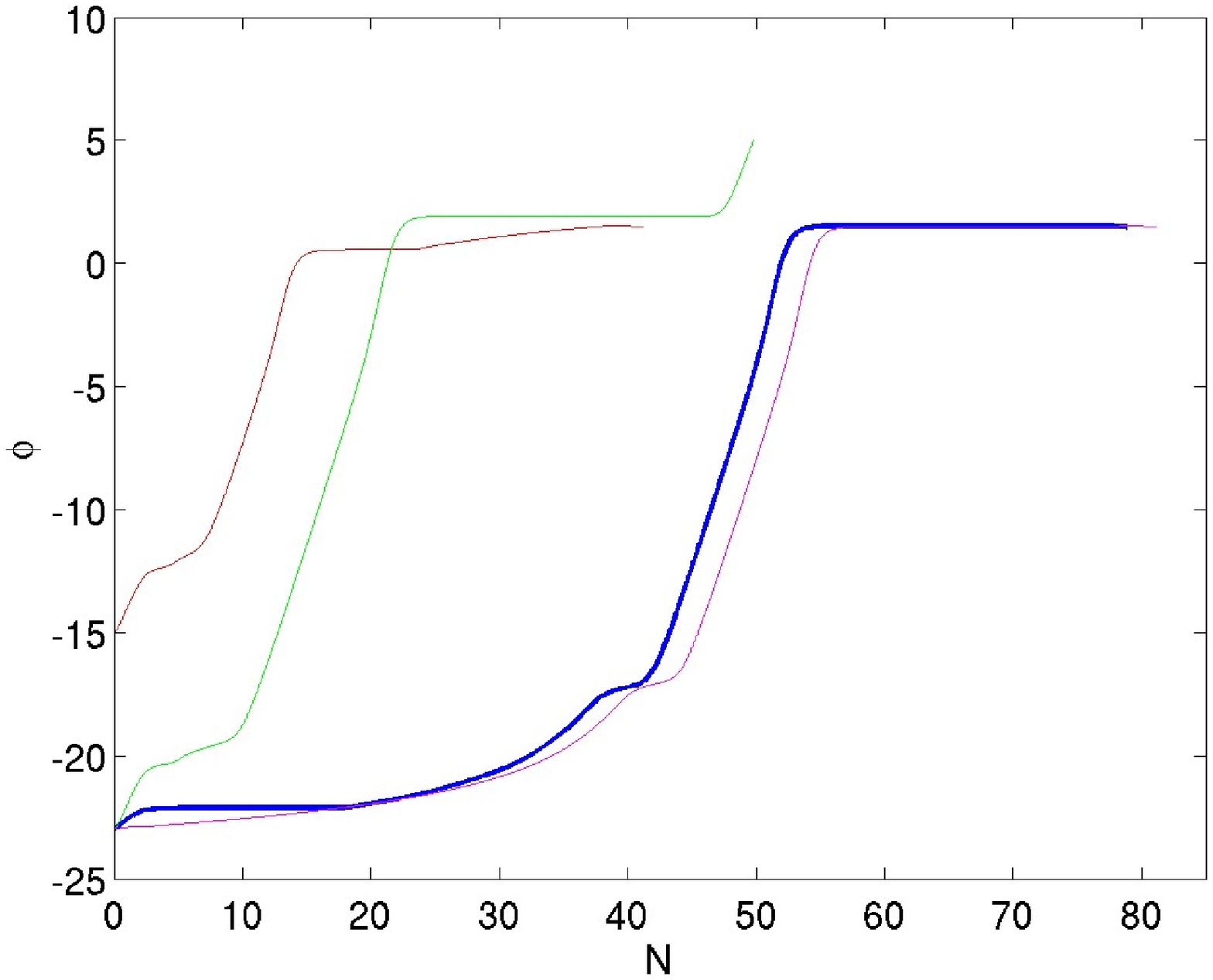}}\label{fig-real_trajs_a}}
   \subfigure[]{
\scalebox{0.37}{\includegraphics{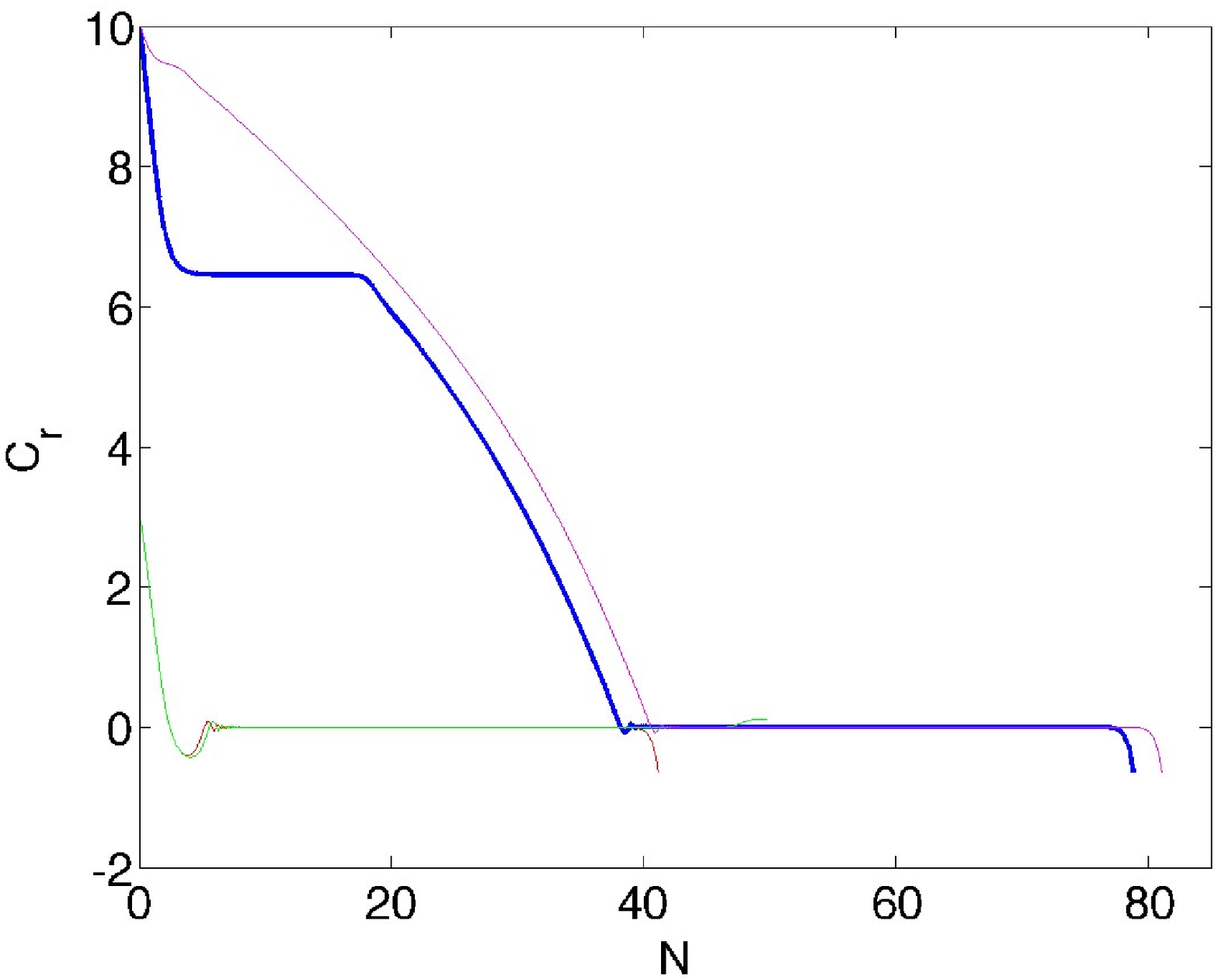}}\label{fig-real_trajs_b}}
  \end{center}
\caption{Various trajectories, each with differing initial conditions 
for $\phi$, $C_r$ and $\Omega_{b0}$. 
Only real fields are considered ($T_i=C_i=0$).
In (a), for $\Omega_{b0}=0.01$, three trajectories are shown (I - solid blue, II - dotted red, III - dashed green).
In (b), for $\Omega_{b0}=0.9$, only one trajectory is shown (IV - dotted magenta).
The greyed region signifies initial 
conditions for trajectories which do not lead to stabilisation. The horizontal lines and regions i-iv are discussed in the text.
In (c) and (d), the evolution of the fields are shown with respect to the $e$-folding number, $N$, for these four trajectories.}
\label{fig-real_trajs}
\end{figure}

We will explain the dynamics using several representative trajectories for which $C_r\ne0$
shown in Fig.~\ref{fig-real_trajs}:
\begin{enumerate}
\item[I] $\phi(0)=-23, \qquad C_r(0)=10, \qquad \Omega_b(0)=0.01 \qquad $ [solid line (blue)]
\item[II] $\phi(0)=-15,\qquad C_r(0)=3, \qquad \Omega_b(0)=0.01 \qquad$ [dotted line (red)]
\item[III] $\phi(0)=-23, \qquad C_r(0)=3 , \qquad \Omega_b(0)=0.01 \qquad$ [dashed line (green)]
\item[IV] $\phi(0)=-23, \qquad C_r(0)=10, \qquad \Omega_b(0)=0.9 \qquad$ [dotted line (magenta)]
\end{enumerate}
We take $\gamma=1$ (a background matter fluid) and consider two cases
$\Omega_{b0}=0.01$ and  $\Omega_{b0}=0.9$.
The initial velocities of the fields are zero.

\begin{figure}[t!]   
  \begin{center}
    \subfigure[]{
\scalebox{0.37}{\includegraphics{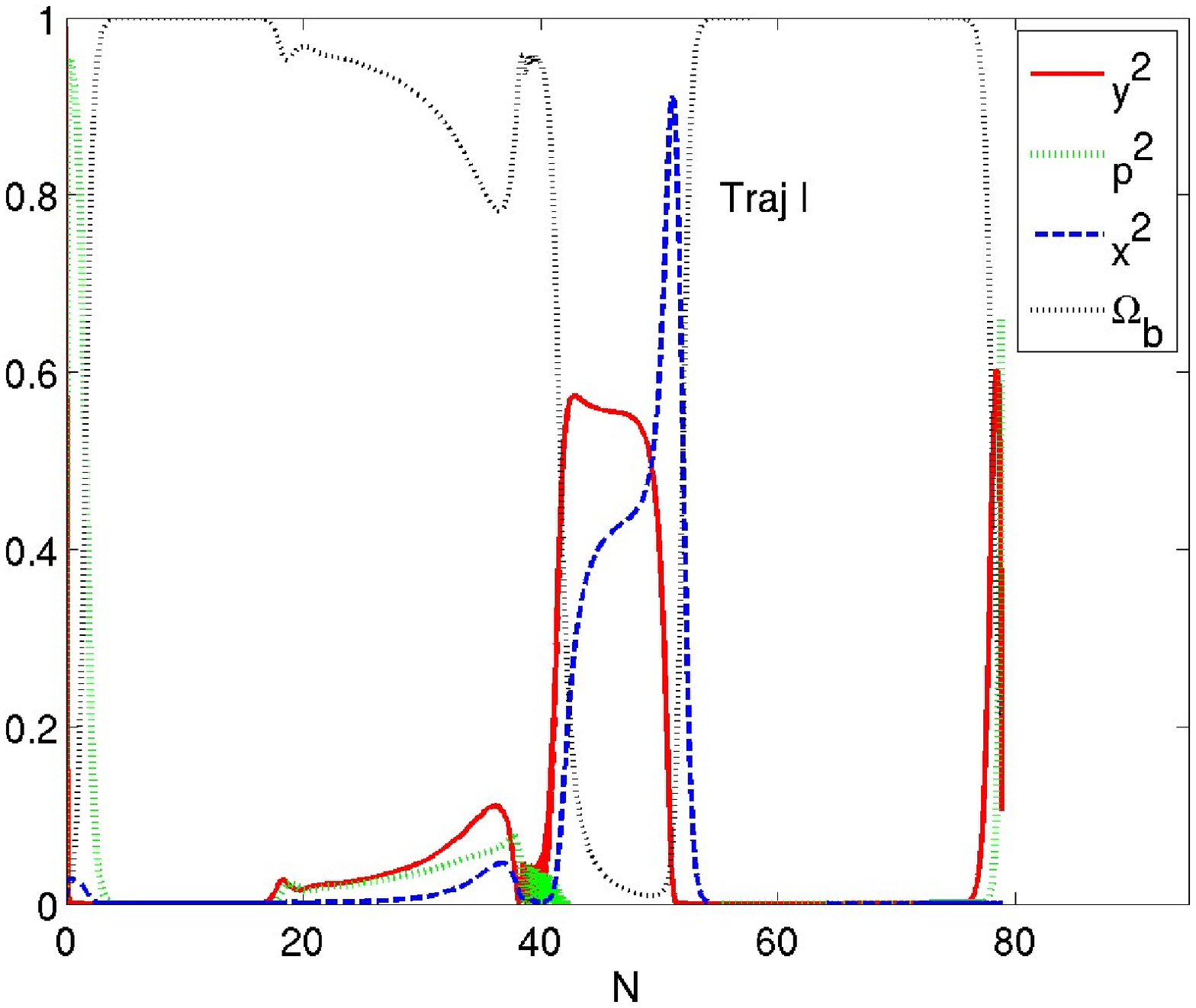}}\label{fig-real_energies1_a}}
    \subfigure[]{
\scalebox{0.37}{\includegraphics{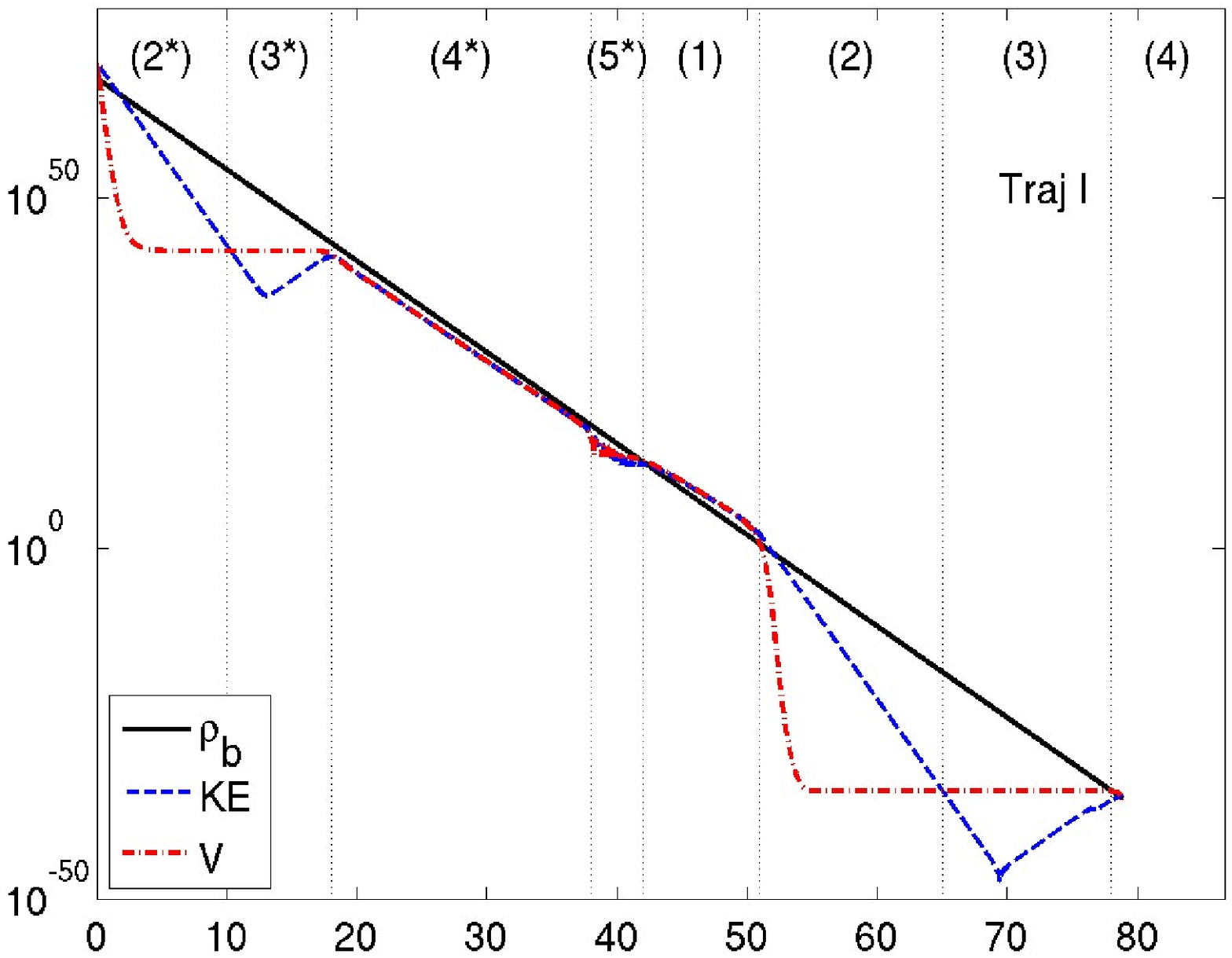}}\label{fig-real_energies1_b}}
  \end{center}
\caption{Breakdown of energy for Trajectory I with initial conditions $\Omega_{b0}=0.01$, $\phi_0=-23$, $C_{r0}=10$.  We consider only the real parts of the fields ($T_i=C_i=0$).}
\label{fig-real_energies1}
\end{figure}

\begin{figure}[t!]
  \begin{center}
    \subfigure[]{
\scalebox{0.37}{\includegraphics{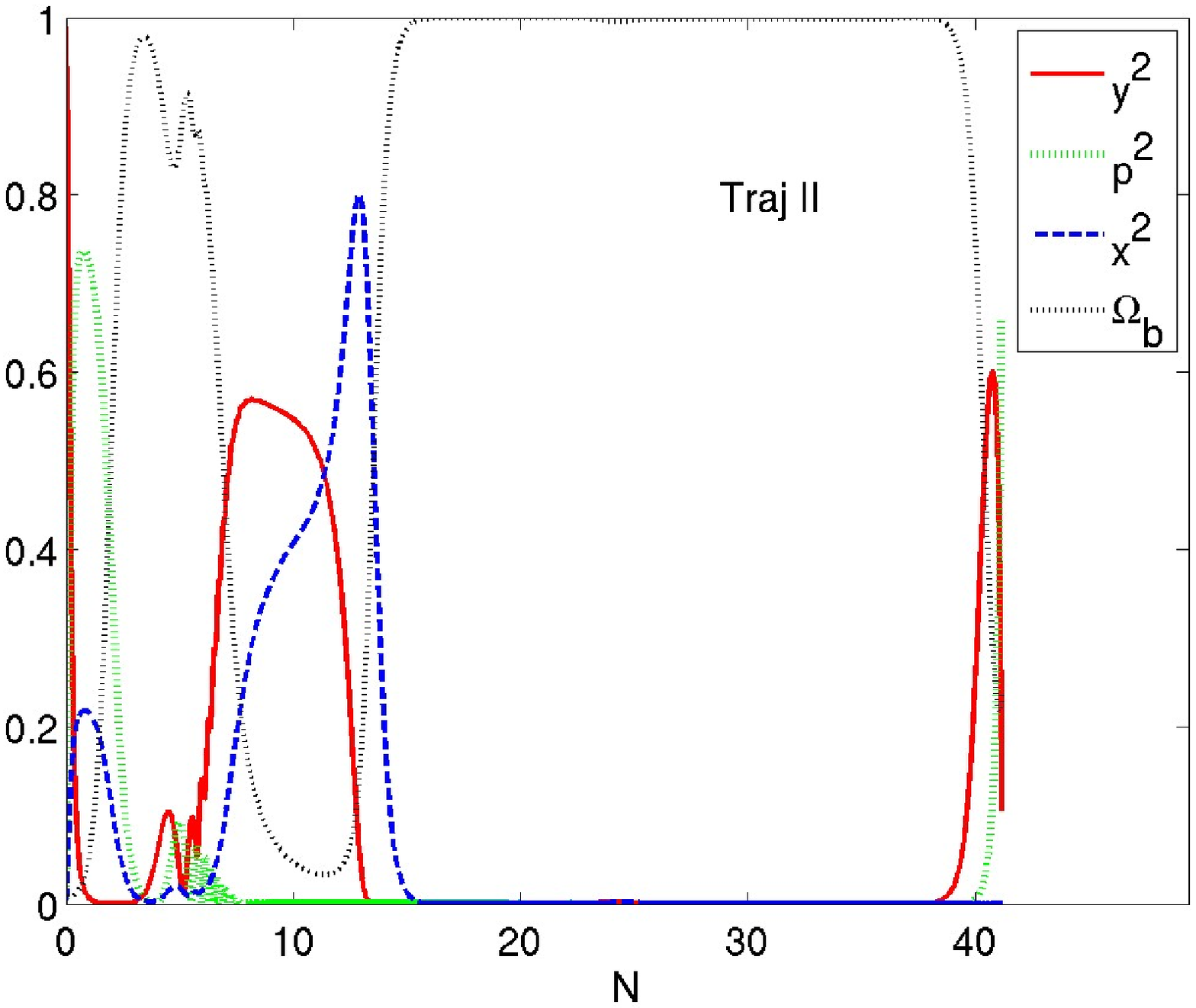}}\label{fig-real_energies2_a}}
    \subfigure[]{
\scalebox{0.37}{\includegraphics{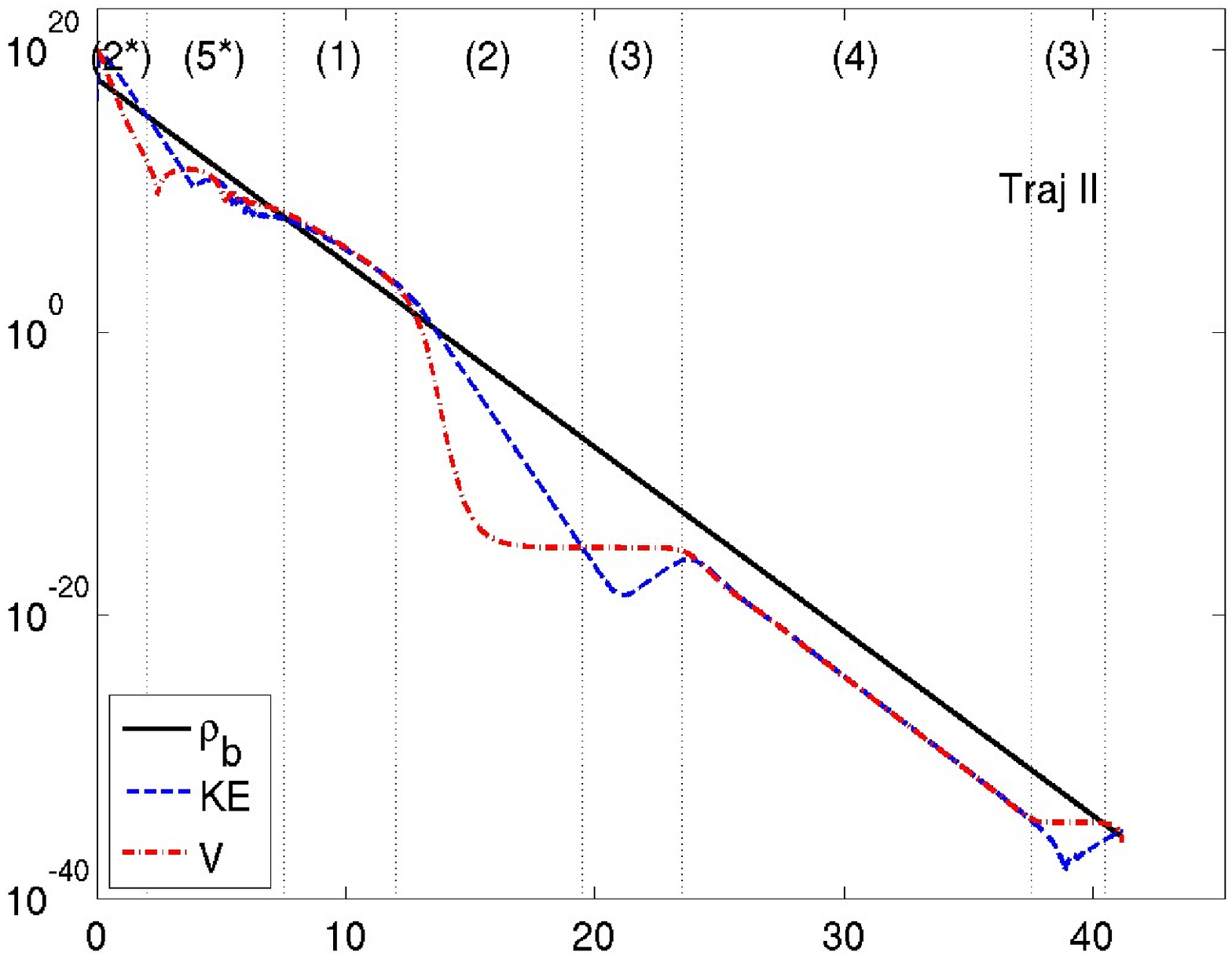}}\label{fig-real_energies2_b}}
  \end{center}
\caption{Breakdown of energy for Trajectory II with initial conditions $\Omega_{b0}=0.01$, $\phi_0=-15$, $C_{r0}=3$.  We consider only the real parts of the fields ($T_i=C_i=0$).}
\label{fig-real_energies2}
\end{figure}

\begin{figure}[t!]
  \begin{center}
     \subfigure[]{
\scalebox{0.37}{\includegraphics{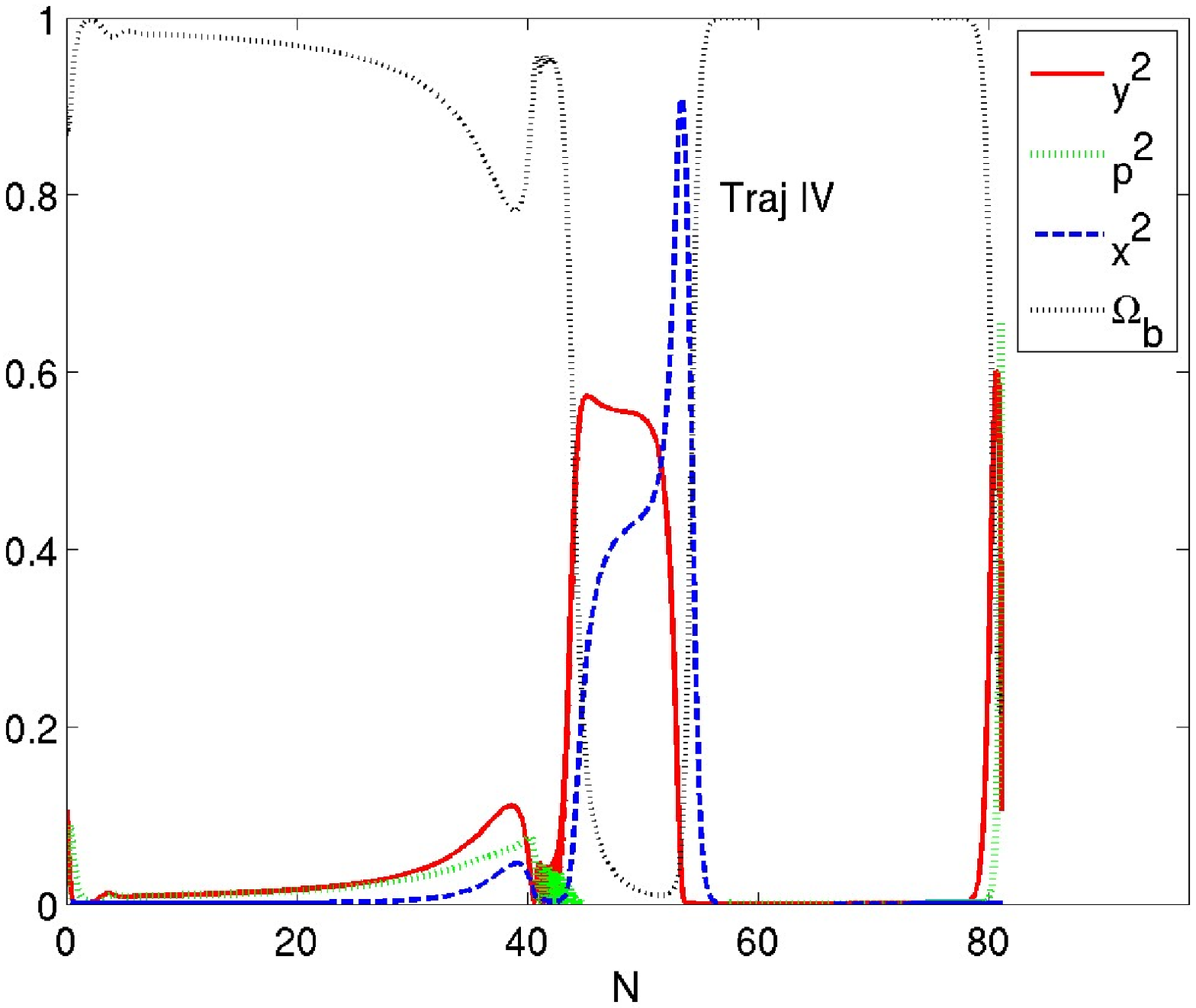}}\label{fig-real_energies3_a}}
    \subfigure[]{
\scalebox{0.37}{\includegraphics{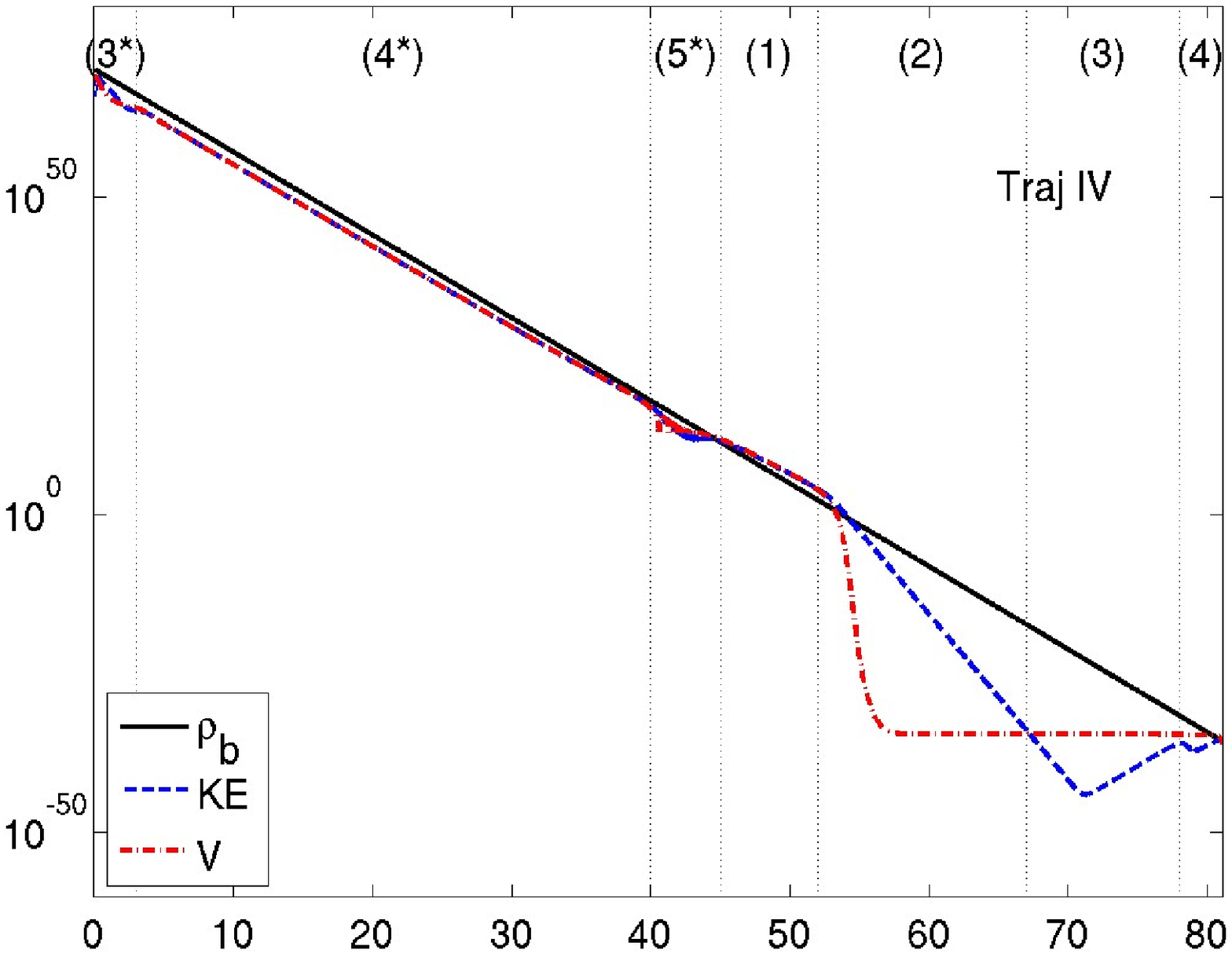}}\label{fig-real_energies3_b}}
  \end{center}
\caption{Breakdown of energy for Trajectory IV with initial conditions $\Omega_{b0}=0.9$, $\phi_0=-23$, $C_{r0}=10$.  We consider only the real parts of the fields ($T_i=C_i=0$).}
\label{fig-real_energies3}
\end{figure}

Fig.~\ref{fig-real_trajs} shows the evolution 
of the fields $\phi$ and $C_r$ against $e$-folding number, $N$, and the 
trajectories in the $\phi$-$C_r$ plane.
The evolution of each energy component
for trajectories I and II are shown in Figs.~\ref{fig-real_energies1} and \ref{fig-real_energies2}.
Trajectory III is identical to Trajectory II (although stabilisation does not occur) and is not shown.  
The evolution of the energy components for trajectory IV is shown in Figs.~\ref{fig-real_energies3}.
Note that Fig.~\ref{fig-real_energies1_b} strongly resembles Fig.~2 
of~\cite{Barreiro:2005ua}. 

As in the single field case,  similar regimes can be identified: (1*) field domination,
(2*) kination, (3*) freeze-out and (4*) scaling.  
The asterix (*) denotes two-field evolution, in order to separate it from the single field case.
When $C_r$ approaches zero, however, these regimes are 
followed by a period of oscillation around $C_r=0$ (Region 5*), further followed by the regimes (1-4) of single field evolution (as in the previous section).
After this, the field is either stabilised or
rolls over.  It should be noted that, due to the steepness in the $C_r$ direction ($\delta>\lambda$), the $C_r$ field arrives  in its 
minimum quickly with the $\phi$ field still running toward the minimum.  Due to this, the second half of the 
dynamics follows single field evolution, as was discussed in the previous section.

\begin{figure}[t!]
\begin{center}
\scalebox{0.7}{\includegraphics{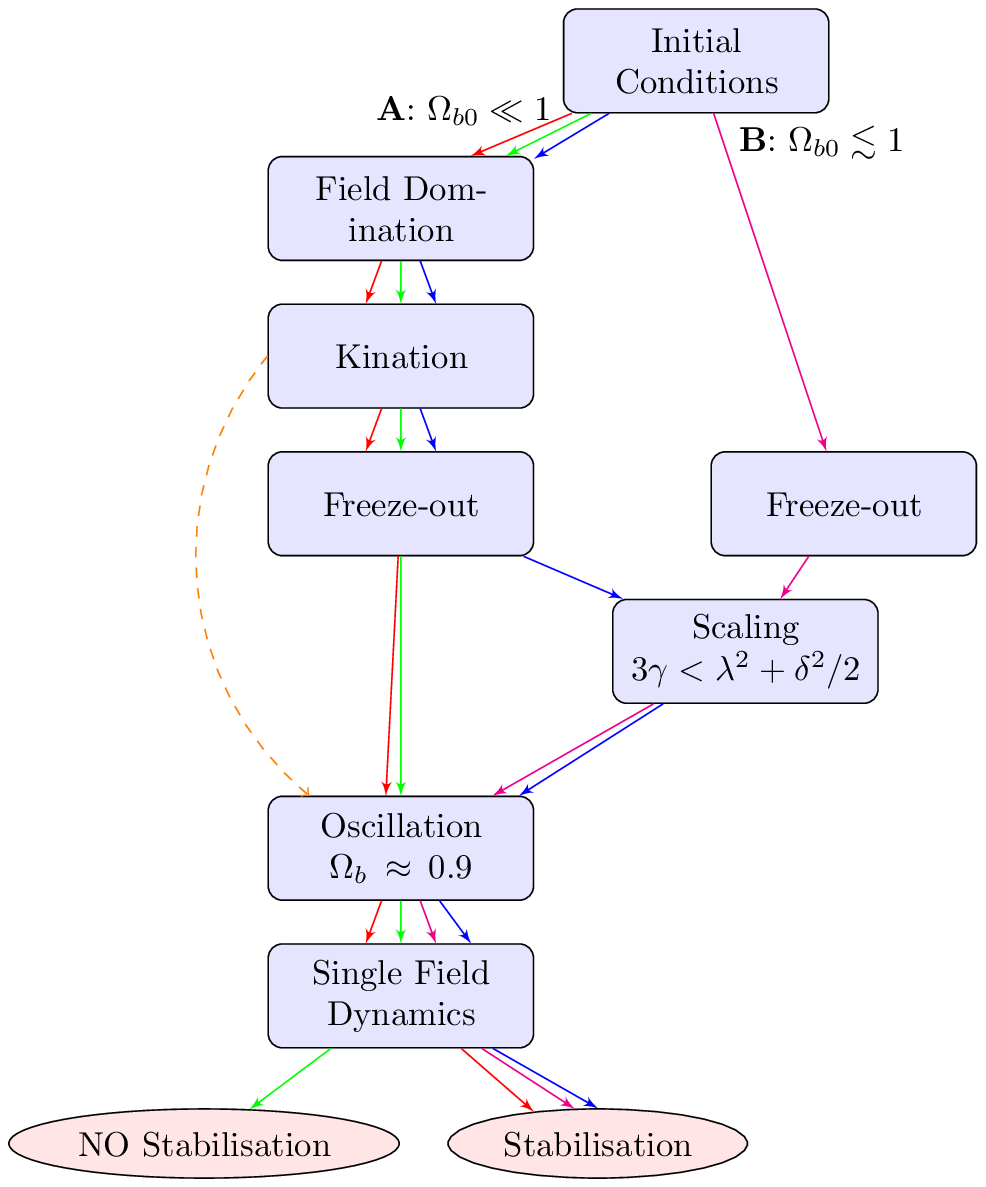}}
\caption{Flowchart of possibilities (Colors: Blue - Trajectory I, Red - Trajectory II, Green -Trajectory III, Magenta - Trajectory IV)}
\label{fig-flow}
\end{center}
\end{figure}

We will now derive some useful approximations for the first part of the evolution, when $|C_r|\ne 0$. 
Depending on the initial value of $\Omega_b$, there are two branches of evolution, A and B.  These are 
shown as a flowchart in Fig.~\ref{fig-flow}, which also shows the path of each example trajectory.

\subsection{Branch A $\Omega_{b0}\ll 1$}

Example trajectories for this branch are seen in Fig.~\ref{fig-real_energies1} and~\ref{fig-real_energies2}, where $\Omega_{b0}=0.01$.  Initially, for a very brief time, the potential energy dominates but since the solution of field domination is not 
stable (see Appendix \ref{AppStability}) this energy is quickly transfered to
the kinetic energy of both fields, $\phi$ and $C_r$.
This change of energy can be described by 
\dis{
x'=\lambda\sqrt{\frac32}y^2,\qquad p'=\delta\frac{\sqrt3}{2}y^2,\qquad 
y'=-\lambda\sqrt{\frac32}xy -\delta\frac{\sqrt3}{2}py,
}
for which the solutions are given by
\dis{
x&=\frac{\sqrt{1-\Omega_{b0}}}{\sqrt{1+\frac{\delta^2}{2\lambda^2}}}\tanh\left[\lambda\sqrt{\frac32}\sqrt{1
+\frac{\delta^2}{2\lambda^2}}N\right],\\
p&=\frac{\sqrt{1-\Omega_{b0}}}{\sqrt{1+\frac{2\lambda^2}{\delta^2}}}\tanh\left[\lambda\sqrt{\frac32}\sqrt{1
+\frac{\delta^2}{2\lambda^2}}N\right],\\
y&=\sqrt{(1-\Omega_{b0})}\sech\left[\lambda\sqrt{\frac32}\sqrt{1+\frac{\delta^2}{2\lambda^2}}N \right].
}
This is the extended solution of Eqns.~(\ref{singlefielddomx}) and~(\ref{singlefielddomy}) for two fields.
The maxima of these solutions, which can be considered to define the start of kination, are 
approximately given by
\dis{
x_{(2)}^2&\simeq\frac{\lambda^2}{\lambda^2+\frac{\delta^2}{2}}(1-\Omega_{b0}),\\
p_{(2)}^2&\simeq\frac{\delta^2/2}{\lambda^2+\frac{\delta^2}{2}}(1-\Omega_{b0}),\\
y_{(2)}^2&\simeq 0.
\label{kination_2field_max}
}
Here we have neglected the evolution of $\Omega_b$, but the error induced should be small.
We shall use the notation that a subscripted bracket will denote the end of the regime of interest, here kination 
(Region 2*), but we omit the asterix for brevity.

During kination, we may assume $y\simeq 0$ and the the solution is
\dis{
x^2&=\frac{x_{(2)}^2}{(1-\Omega_{b{(2)}})+\Omega_{b{(2)}}e^{3(2-\gamma)(N-N_{(2)})}}, \\
p^2&=\frac{p_{(2)}^2}{(1-\Omega_{b{(2)}})+\Omega_{b{(2)}}e^{3(2-\gamma)(N-N_{(2)})}}, \\
\Omega_b&=\frac{\Omega_{b{(2)}}e^{3(2-\gamma)(N-N_{(2)})}}{(1-\Omega_{b{(2)}})+\Omega_{b{(2)}}e^
{3(2-\gamma)(N-N_{(2)})}}.
}
When the potential energy negligible compared to the kinetic energy, from the Klein-Gordon equation
\dis{
\ddot{\phi}+3H\dot{\phi}\simeq0,
}
for which the solution is $\dot{\phi}\propto a^{-3}$.
Therefore the kinetic energy is proportional to $a^{-6}$. This is the same for the $C_r$ field.

Alternatively, in this region, we can obtain
\dis{
\frac{p'}{x'}=\frac{p}{x}=\sqrt2\frac{C_r'}{\phi'}={\rm constant},
}
where we have used the original definition of $x$ and $p$ for the third term.
This gives a direct relation between $x$ and $p$
\dis{
p=\frac{p_{(2)}}{x_{(2)}}x,\qquad  C_r-C_{r_{(2)}}=\frac{p_{(2)}}{\sqrt2x_{(2)}}(\phi-\phi_{(2)}),
}
where we have used the initial values of kination from Eqn.~(\ref{kination_2field_max}).
The freeze-out point can be calculated by integrating these solutions
\dis{
\Delta\phi&=\frac{x_{(2)}}{\sqrt{1-\Omega_{b{(2)}}}}\frac{2\sqrt6}{3(2-\gamma)}
\left[\sinh^{-1} \left(\sqrt{\frac{1-\Omega_{b{(2)}}}{\Omega_{b{(2)}}}} \right)
-\sinh^{-1}  \left(\sqrt{\frac{1-\Omega_{b{(2)}}}{\Omega_{b{(2)}}}}e^{-3(2-\gamma)N/2} \right) \right],
\\
\Delta C_r&=\frac{p_{(2)}}{\sqrt{1-\Omega_{b{(2)}}}}\frac{2\sqrt3}{3(2-\gamma)}
\left[\sinh^{-1} \left(\sqrt{\frac{1-\Omega_{b{(2)}}}{\Omega_{b{(2)}}}} \right)
-\sinh^{-1}  \left(\sqrt{\frac{1-\Omega_{b{(2)}}}{\Omega_{b{(2)}}}}e^{-3(2-\gamma)N/2} \right) \right].
\label{C_r3}
}
where $\Delta\phi=\phi_{(3)}-\phi_{(2)}$ and $\Delta C_r=C_{r(3)}-C_{r(2)}$
Note however, that when the initial background density is small enough, 
the transfer of potential energy to kinetic energy is so quick
that we can consider starting at a point $\phi_0$, $C_{r0}$ with kinetic energies given by $x_{(2)}$,$p_{(2)}$.

\subsection{Branch B $\Omega_{b0}\lesssim 1$}

An typical trajectory of this branch is shown in Fig.~\ref{fig-real_energies3}.  If $\Omega_b$ is 
initially large, then the potential energy can never dominate and kination is not achieved. 
Instead, the equation of motion for the background fluid is given by
\dis{
\Omega^{\prime}_b=-3\Omega_b\left[(\gamma-2)(1-\Omega_b)+2y^2 \right].
}
As potential energy is converted to kinetic energy, $y^2$ becomes negligible in a few e-folds and we 
quickly obtain the solution
\dis{
\Omega_b=\left(1+\frac{1-\Omega_{b0}}{\Omega_{b0}}e^{3(\gamma-2)N} \right)^{-1}.
}
Though the fields obtain some kinetic energy from the potential energy, it is 
suppressed compared to the background density.
Quickly, $\Omega_b\rightarrow 1$ and the fields become frozen.
This process is very fast and we can make the approximation $\phi_{(3)}\approx\phi_0$ and $C_{r(3)}
\approx C_{r0}$

\subsection{Scaling}
\label{subsecScaling}
We calculate the full scaling solution and condition in Appendix~\ref{App1}.  We will use the main results 
in this section.
The two-field scaling condition is given by
{\begin{eqnarray}
3\gamma<\lambda^2+\delta^2/2.
\label{two_field_scaling_cond}
\end{eqnarray}
At the freeze-out point, if the condition for scaling is satisfied,
then the fields track the background fluid.
We define 
\begin{eqnarray}
\Gamma_{ij} = \frac{V V_{ij}}{V_i V_j}.
\end{eqnarray}
Making the assumption that
\dis{
\Gamma_{\phi\phi}\simeq\Gamma_{C_r C_r}\simeq \Gamma_{\phi C_r}\simeq 1,
}
which is reasonable in the ranges $\phi\lesssim -8$ and
 $|C_r|\gtrsim 1$ for our potential,
we obtain 
\dis{
x_c=\sqrt{\frac{3}{2}}
\frac{\tilde\gamma}{\lambda},\qquad p_c=\frac{\sqrt3}{2}\frac{\delta\tilde\gamma}{\lambda^2},
\qquad y_c^2=\frac32\frac{\tilde\gamma}{\lambda^2}\left(2-\tilde\gamma-\frac{\delta^2}{2\lambda^2}
\tilde\gamma \right),
\label{scaling}
}
where 
\dis{
\tilde\gamma=\gamma\left(1+\frac{\delta^2}{2\lambda^2} \right)^{-1}.
}
This is a valid solution only when $\Gamma_{ij}\approx1$.  When $C_r=1$, however, $\delta$ changes form and $\Gamma_{Cr Cr}$ begins to decrease (and later increases again).
This deviation from $\Gamma\approx1$ leads to an increase in $\Omega_b$ (as seen around $N\sim35$ in Fig.~\ref{fig-real_energies1}).
Scaling proceeds until $C_r$ goes to around $0$, where the scaling condition
is violated.

During scaling, the background fluid density takes the form
\dis{
\Omega_b=1-x_c^2-p_c^2-y_c^2=1-\frac{3\tilde\gamma}{\lambda^2}
=1-\frac{3\gamma}{\lambda^2+\delta^2/2}
}
and the trajectory is specified by
\dis{
\frac{p_c}{x_c}=\frac{\delta}{\sqrt2 \lambda}=\sqrt2\frac{dC_r}{d\phi}.
}
From Fig.~\ref{lambda+delta}, we may take $\lambda$ approximately constant and from Eqn.~(\ref
{V_real})
\dis{
\delta \approx -2(C_r+\frac{1}{C_r})
}
so that
\dis{
\phi=\phi_{(3)}-\frac{\lambda}{2}\log\frac{1+C_r^2}{1+C_{r(3)}^2}.
\label{phiendscal}
}

At the end of the scaling regime, as found numerically,
the fractional energy of a matter background fluid is 
\dis{
\Omega_b\simeq 0.93,
\label{omega_93}
}
and the kinetic energy of $\phi$ is small.
We find that this fractional background density is numerically independent of the initial values
of the dynamics and is determined only by the existence of scaling solution (i.e. potential dependent).
Additionally, from Eqn.~(\ref{phiendscal}), when the trajectory reaches $C_r=0$, we find
\dis{
\phi_{(5)}\approx\phi_{(3)}-\frac{\lambda}{2}\log\frac{1}{1+C_{r(3)}^2},
}
where either [$\phi_{(3)}$,$C_{r(3)}$] are calculated from Eqn.~(\ref{C_r3}) (Branch A) or [$\phi_0$,$C_{r0}$] 
(Branch B).
For a matter background and $\Omega_{b0}\approx 1$, trajectories always scale.  When $\Omega_{b0}\ll 1$,  only trajectories with $\vert C_{r0}\vert\gtrsim 4$ have time to enter a scaling regime.  This region is depicted by regions (i) in Fig.~\ref{fig-real_trajs_c} and separated with a dotted line.

\subsection{Oscillation}

If the trajectory undergoes scaling, then it is clear from Eqn.~(\ref{scaling}) that, for our potential at least, $p_c>x_c$ and there is more kinetic energy in $C_r$ than in $\phi$.
When the trajectory reaches the $C_r$ minimum ($C_r=0$), the trajectory oscillates but $\phi$ is almost constant; it acts as if it is almost frozen at a point $\phi_{f2}$ with $\Omega_b\approx 0.93$.

If the trajectory does not scale (like Trajectories II and III), the trajectory gently oscillates around the minimum, with $\phi$ not fixed. 
Due to the relative gradients, $\lambda$ and $\delta$, the kinetic energy is shared between the fields, but the expansion of the universe acts to damp the oscillations.  Finally, the background fluid dominates and the fields stop rolling at a point $\phi_{f2}$.
At this point, we find numerically that $\Omega_b\approx 0.9$ once again.

After freeze-out, $C_r\approx 0$ and the trajectory enters single field dynamics as discussed in Section \ref{sec_review_single}.

\subsection{Stabilisation Regions}
We shall characterise a stabilising trajectory by the initial conditions for $\phi$ and $C_r$ which lead to 
final stabilisation.  
Samples of the stabilisation regions for 
two real fields can be seen in Fig.~\ref{fig-real_trajs}.  Greyed out regions denote initial field values for 
trajectories which do not result in stabilisation.
We will see that the important factor in stabilisation is the point $\phi_{f2}$, the value at which the $\phi$-
field freezes on the $C_r=0$ line, after either scaling or oscillation.  

\subsubsection{$\Omega_{b0}\lesssim 1$}
We will first explain the stabilisation region for a large background fluid density, such as when $
\Omega_{b0}=0.9$ as shown in Fig.~\ref{fig-real_trajs_d}.
A trajectory starting at large $\vert C_r\vert$ will follow similar dynamics as Trajectory IV.  After freezing 
at a point [$\phi_{f1}$,$C_{r(f1)}$], the fields will quickly reach a scaling regime.  
When $C_r\approx0$, 
scaling ends and the $C_r$-field oscillates around the ``valley" and $\phi$ is effectively fixed at $\phi_{f2}$.
Henceforth, the trajectory follows single field dynamics, with initial conditions $\phi_{f2}$ and $\Omega_{b}\approx 0.93$.

As discussed in Section \ref{sec_review_single}, stabilisation depends merely on these initial conditions.  
Stabilisation may occur only when $-17<\phi_{f2}<1$ for $\Omega_b=0.93$, as is seen in Fig.~\ref{fig-real_trajs_d}.  
The slight asymmetry of the stabilisation region (seen for large $\phi$) is due to the asymmetry of the potential;
the local minimum does not occur at $C_r=0$.  Trajectories which start close to the minimum are easier to 
stabilise.

We have numerically simulated trajectories only within $\vert C_r\vert<10$.  Note however, that all 
trajectories with $\phi_{f2}$ within the stabilisation bounds at $C_r=0$ will lead to stabilisation.  Therefore 
the allowed $C_r$ region extends to $|C_r|\rightarrow\infty$, due 
to the scaling solution.

\subsubsection{$\Omega_{b0}\ll1$}
The stabilisation region for $\Omega_{b0}\ll 1$ looks vastly different from the previous example, as can 
be seen in Fig.~\ref{fig-real_trajs_d}.
However, the underlying structure is the same.  In regions (i), similar constraints are seen.  These are 
due to scaling solutions, which result in the $\phi$-field freezing on the $C_r$ axis between $-17<\phi_{f2}<1$ as before.

In region (ii), however, though the scaling condition is satisfied, 
the field does not have enough time to reach this scaling regime before reaching
$C_r=0$. Thus the field has enough kinetic energy at the crossing point to oscillate around $C_r=0$ (seen in Fig.~\ref{fig-real_trajs_d}). 
As described earlier, the trajectory settles into the minimum at $C_r=0$ at a point $\phi_{f2}$.
As long as $\phi_{f2}$  is within the range $-17<\phi_{f2}<1$, the trajectory will lead to stabilisation. Therefore Trajectory II stabilises ($\phi_{f2}\approx -12$) whilst Trajectory III does not ($\phi_{f2}\approx -20$).

The conical boundary of region (ii) depends on the energy density of the background fluid.
For smaller $\Omega_{b0}$, the kinetic energies are less damped and oscillate more widely.
Since the field gains more kinetic energy initially (from potential energy) and is less damped, the fields travel further before settling at a point $\phi_{f2}$.
Therefore as $\Omega_{b0}$ decreases, the conical bound moves left and the stabilisation area increases.

The asymmetry seen between regions (iii) and (iv) is once more again to the asymmetry of the potential;  trajectories starting in region (iii) are pushed towards the minimum (due to the gradient) and stabilise, whereas trajectories from region (iv) are pushed away from the minimum.

\section{Dynamics of Complex Fields}
\label{secComplex}
In this section, we include the dynamics of the imaginary parts of the fields, $C_i$ and $T_i$, in order to fully generalise our model.
The existence of non zero imaginary parts changes the shape
of the potential in the $\phi$-$C_r$ plane during evolution.
As an example we consider a trajectory with initial values:
\dis{
\phi=-2,\qquad C_r= 0.25,\qquad T_i=0,\qquad C_i=1,\qquad \Omega_{b0}=0.9.
\label{imag_init}
}
The evolution of the fields for this trajectory are shown in Fig.~\ref{fig-imag_overview}.  It can be seen that, even though $T_{i0}=0$, this field is given a ``kick'' since $C_{i}\ne 0$
\begin{figure}[t!]
  \begin{center}
  \begin{tabular}{c c}
    \includegraphics[width=0.4\textwidth]{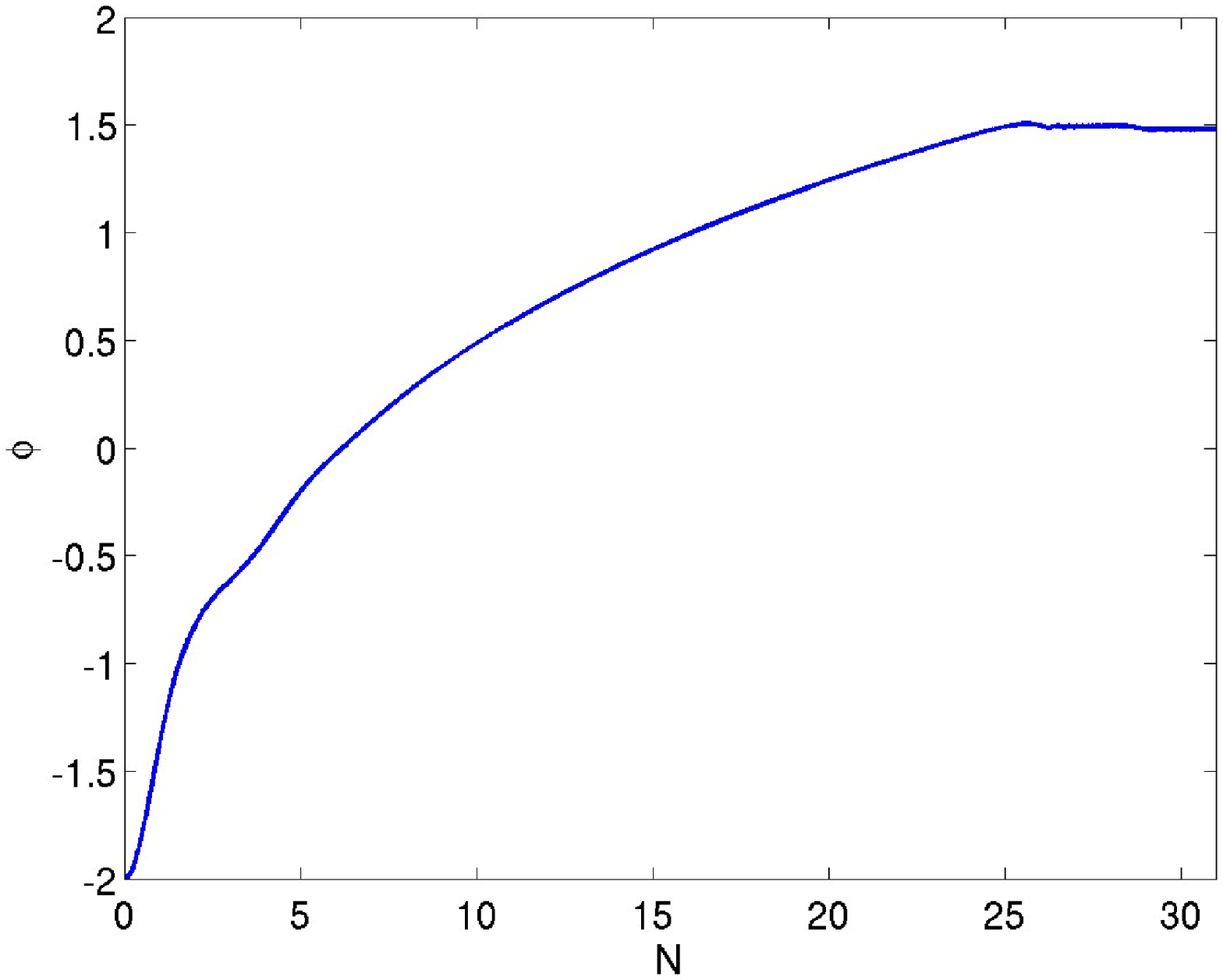}
&
    \includegraphics[width=0.4\textwidth]{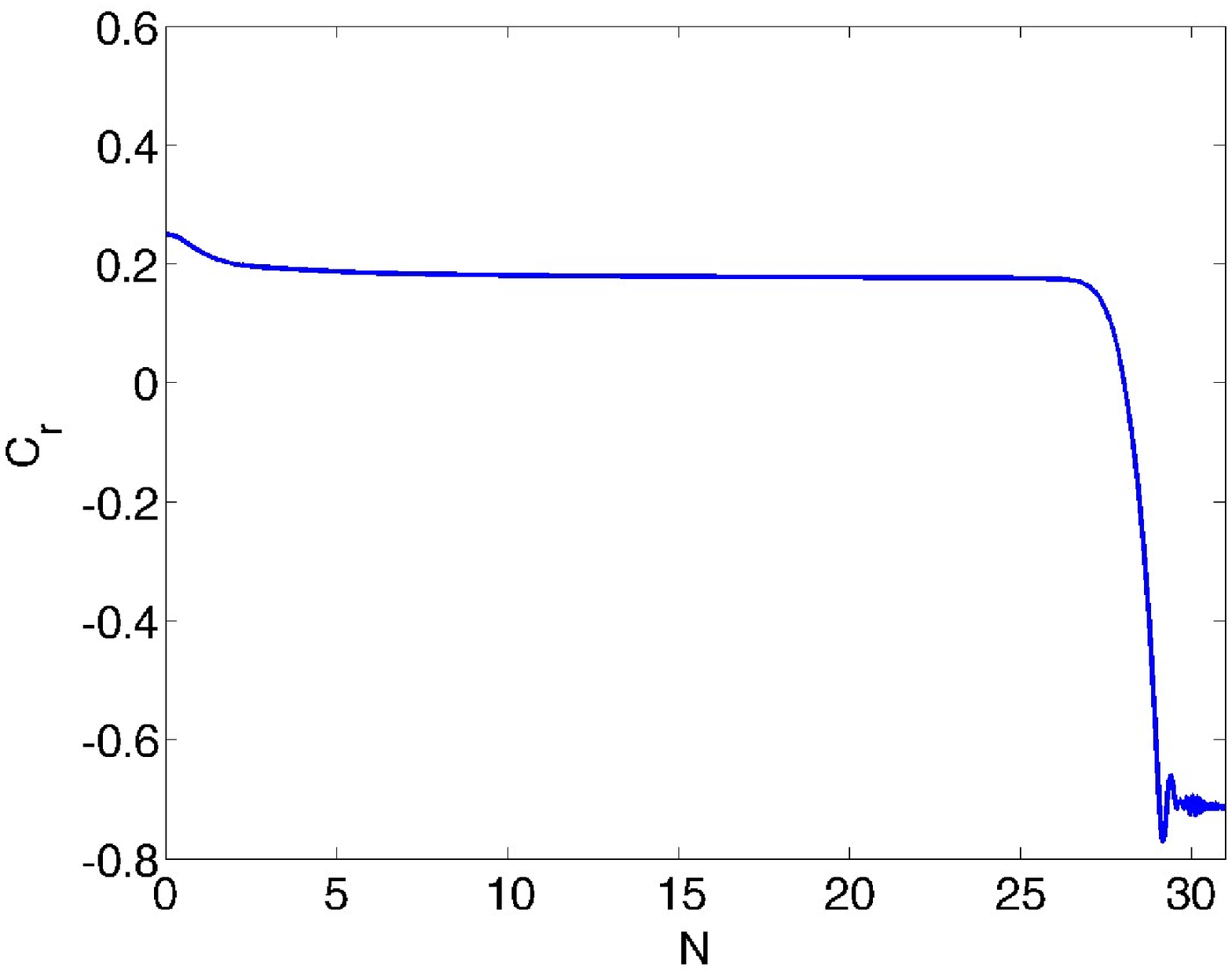}
\\
    \includegraphics[width=0.4\textwidth]{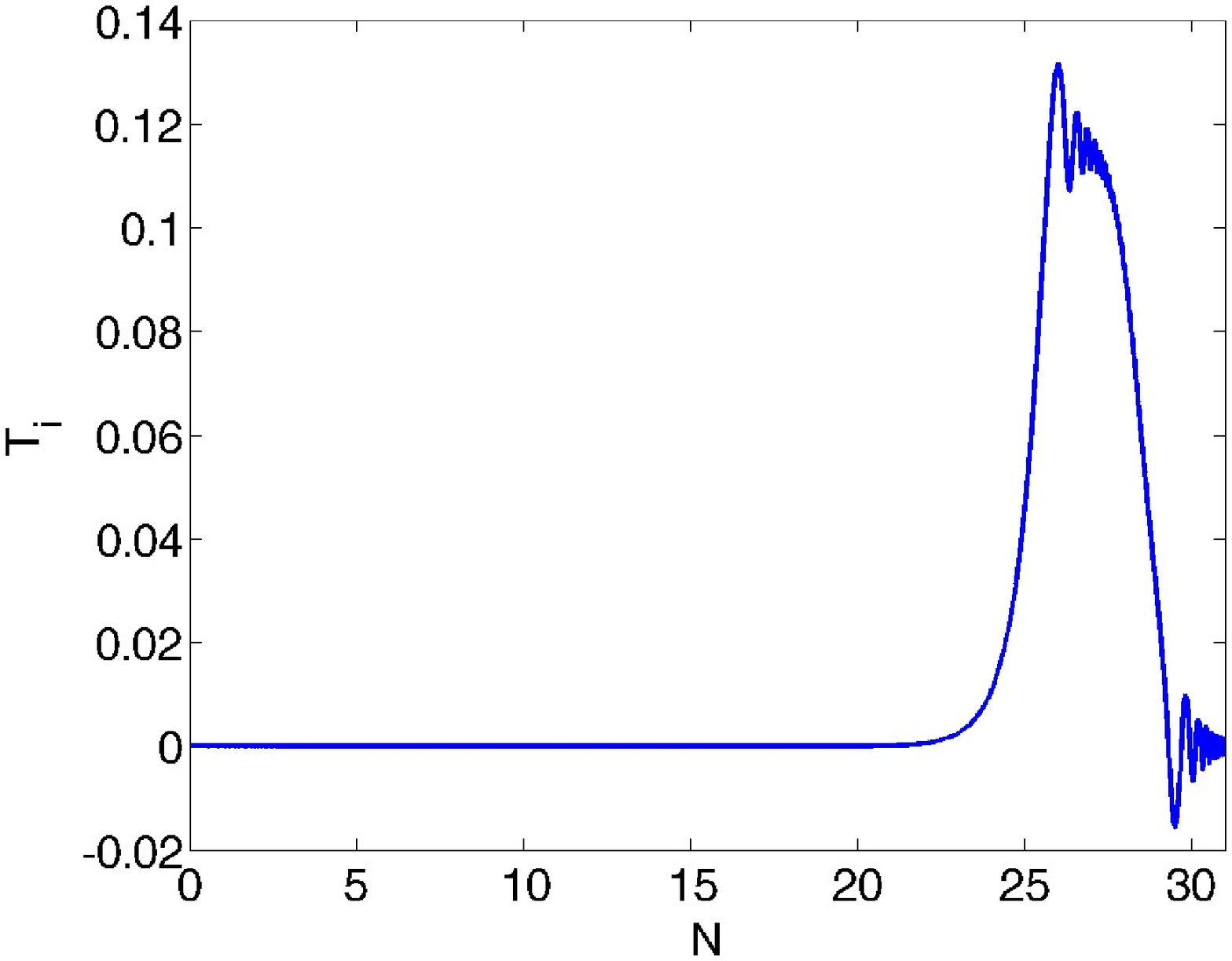}
&
    \includegraphics[width=0.4\textwidth]{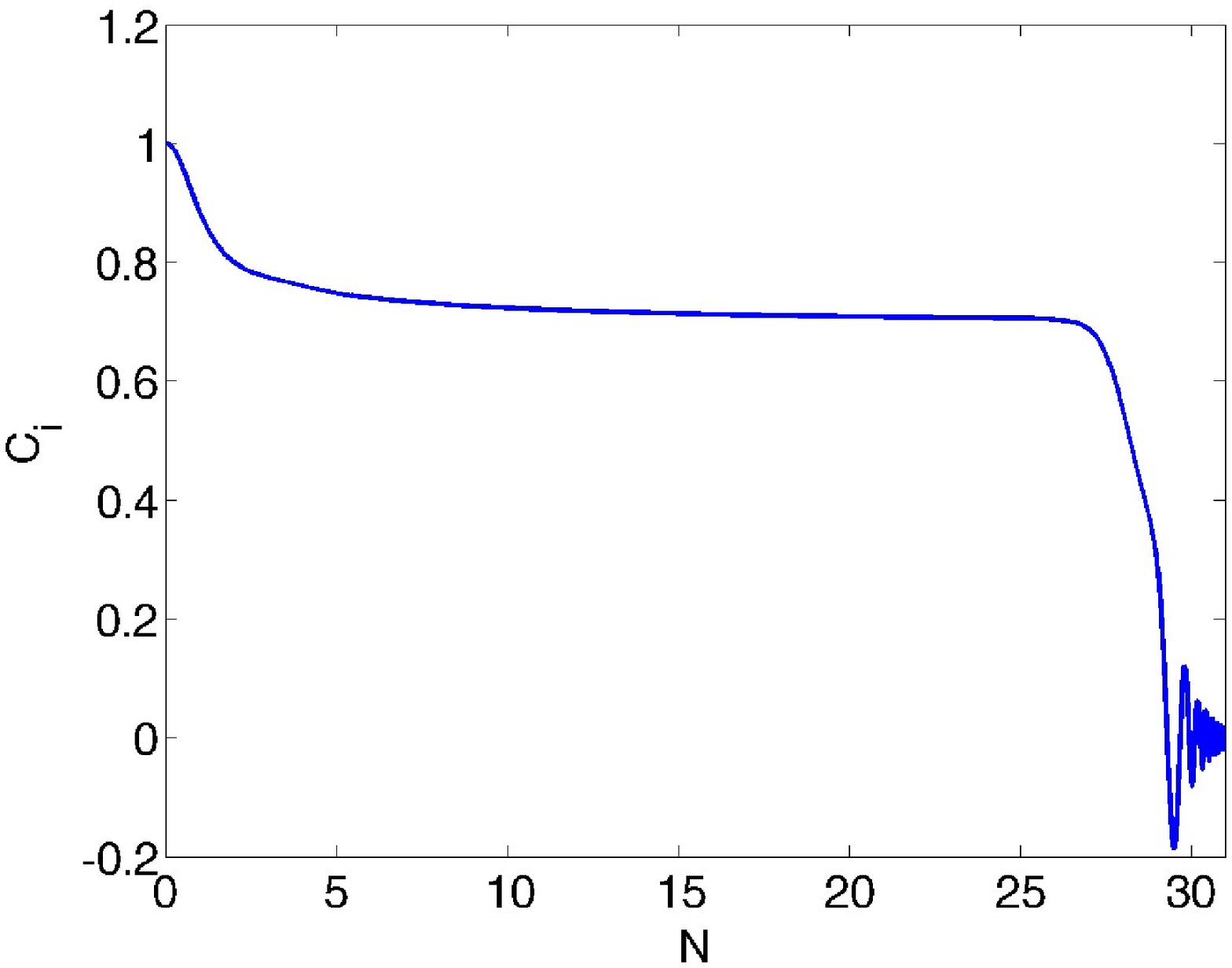}
    \end{tabular}
  \end{center}
\caption{An example trajectory with evolving $\sigma_i$ and $C_i$, when $\Omega_{b0}=0.9$.  The initial conditions are given in Eqn. (\ref{imag_init}).}
\label{fig-imag_overview}
\end{figure}
We show the evolution of each energy component in Fig.~\ref{fig-imag_energies}, where the kinetic energy of $T_i$ is seen to be negligible.  This is due to the fact that $T_{i0}=0$ and is not generic.
\begin{figure}[t!]
  \begin{center}
    \subfigure[]{
\scalebox{0.37}{\includegraphics{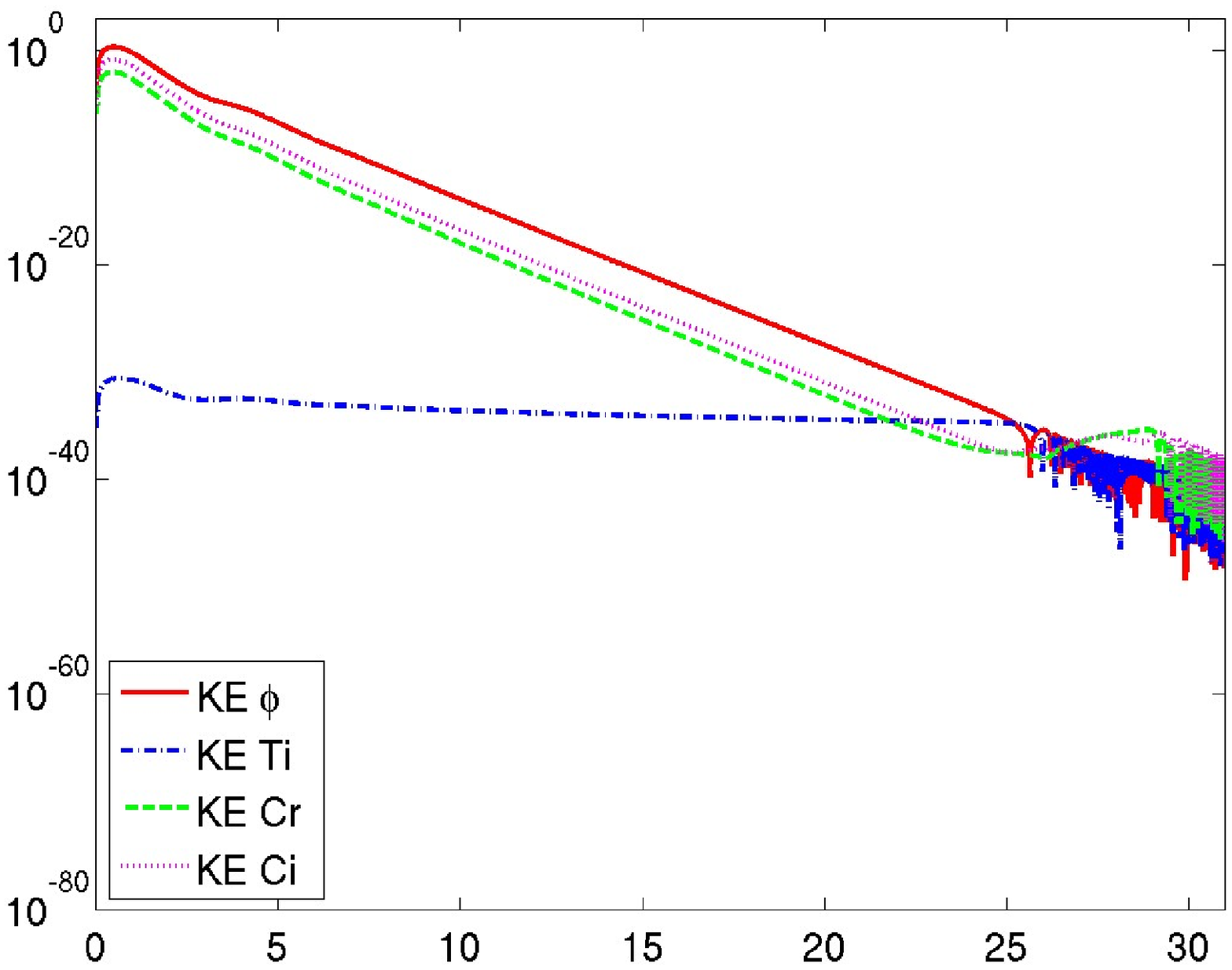}}\label{fig-imag_energies_a}}
    \subfigure[]{
\scalebox{0.37}{\includegraphics{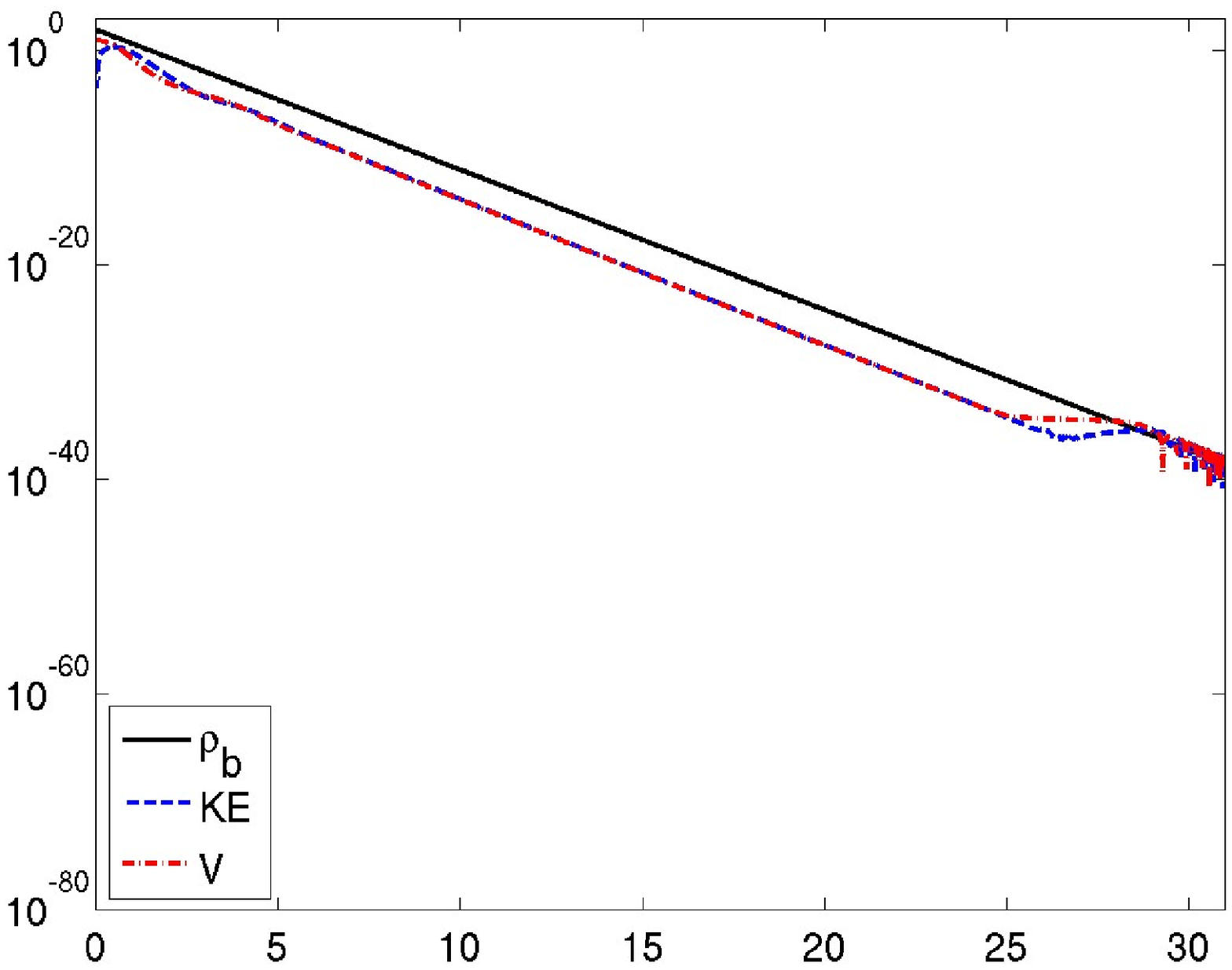}}\label{fig-imag_energies_b}}
  \end{center}
\caption{The evolution of each energy component for the trajectory given by the initial conditions in Eqn. (\ref{imag_init}).  (a) The breakdown of kinetic energies. (b) The total breakdown.  Note that the kinetic energy of $T_i$ is negligible since $T_{i0}=0$; it is not a generic result. }
\label{fig-imag_energies}
\end{figure}
\begin{figure}[t!]   
  \begin{center}
    \includegraphics[width=1.0\textwidth]{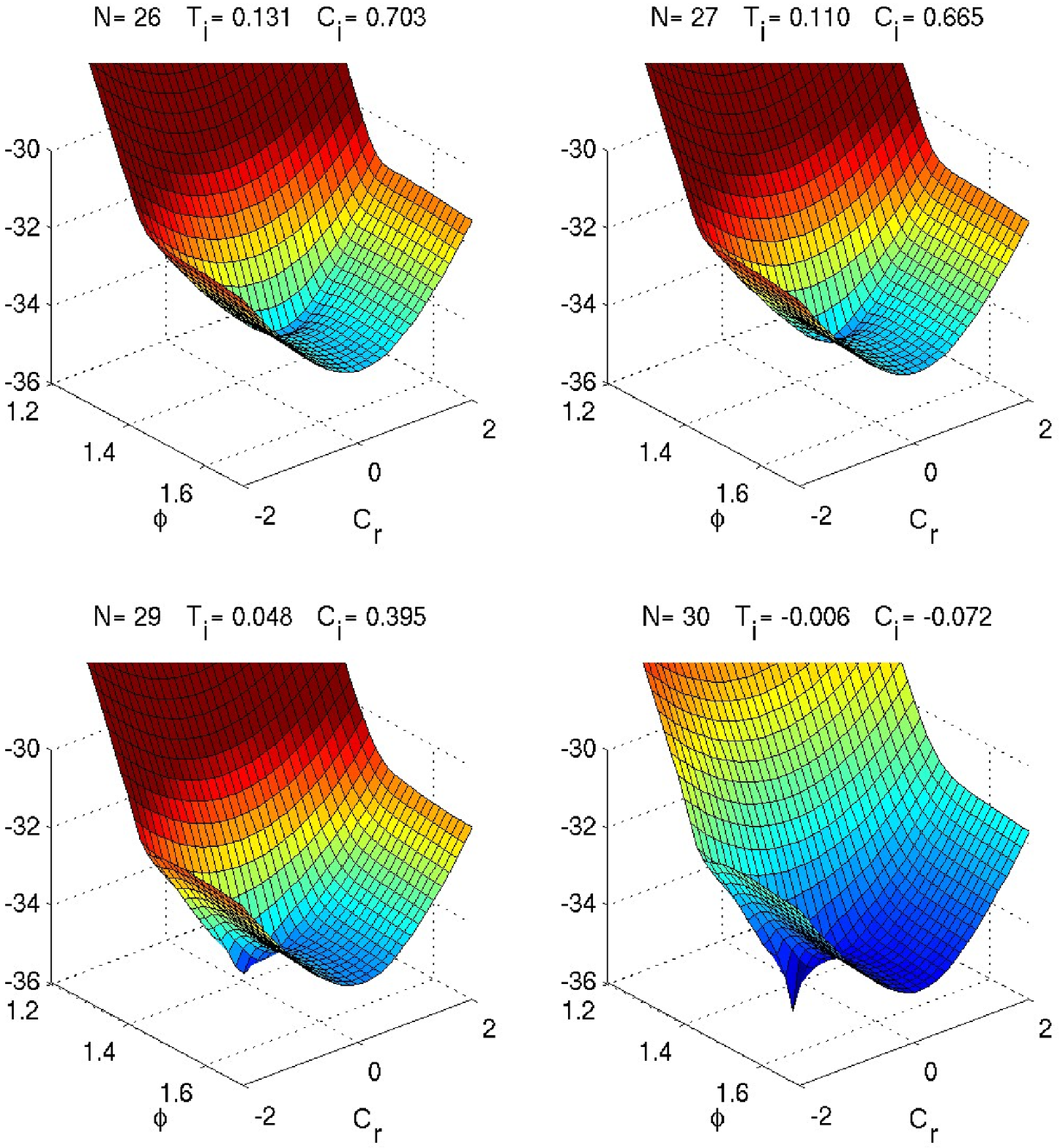}
  \end{center}
\caption{The potential at different efolding values ($N=26,27,29,30$), for two complex fields ($T_i$ and $C_i$ are non-zero), showing how the existence of the
minimum (in the $\phi$-$C_r$ plane) depends on the imaginary field values. The initial conditions are given in Eqn. (\ref{imag_init}).  At $N\approx 26$, the local minimum has almost disappeared since $C_i$ is large, but reappears at later times when $C_i\approx 0$.  The evolution of the potential works to nudge the fields into the local minimum.}
\label{fig-imag_pot}
\end{figure}
In Fig.~\ref{fig-imag_pot} we show the contour plots of the potential in the $\phi-C_r$ plane
for the above trajectory at e-folding numbers $N=26,27,29,30$. 
In the top lefthand plot, where $T_i \simeq 0.1$ and $C_i\simeq 0.7$, the local
minimum almost disappears. However, at larger N and smaller $C_i$, the potential changes shape and
finally around $C_i\simeq 0$ the local minimum re-appears and the fields
find this minimum. Thus we see that evolution of the imaginary fields
can change the possibility of stabilisation. 
In fact, the evolution of the potential works to push the fields into the local minimum; as the fields get close to the position of the local minimum, the potential barrier effectively increases (the plateau is raised) and the fields are pushed back towards the re-emerging minimum.

In order to understand the full stabilisation region, a coarse 4-dimesional grid of trajectories was analysed numerically, in which each point represents the initial conditions for [$\phi$,$T_i$,$C_r$,$C_i$].  Three such grids were run, for $\Omega_{b0}=10^{-10},10^{-2},0.9$.  The results are shown in Fig.~\ref{fig-vol} for the 3-D volume $\phi$-$C_r$-$C_{i}$.  We do not show the $T_{i}$ direction, since all values lead to identical surface plots (this is due to the symmetry seen in Fig.~\ref{fig-ex2}).  
\begin{figure}[t!]
  \begin{center}
  \begin{tabular}{c c}
    \includegraphics[width=0.5\textwidth]{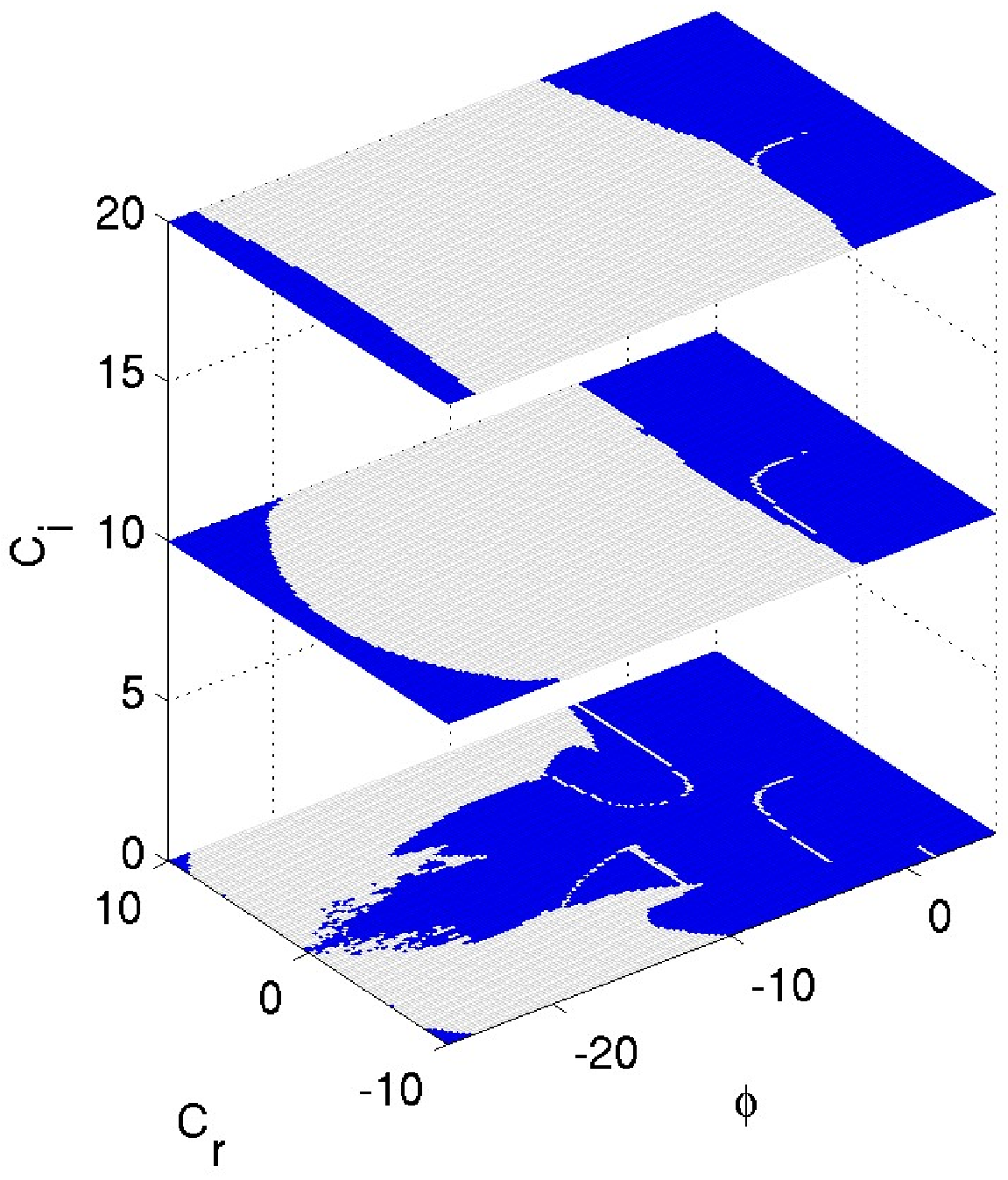}
&
    \includegraphics[width=0.5\textwidth]{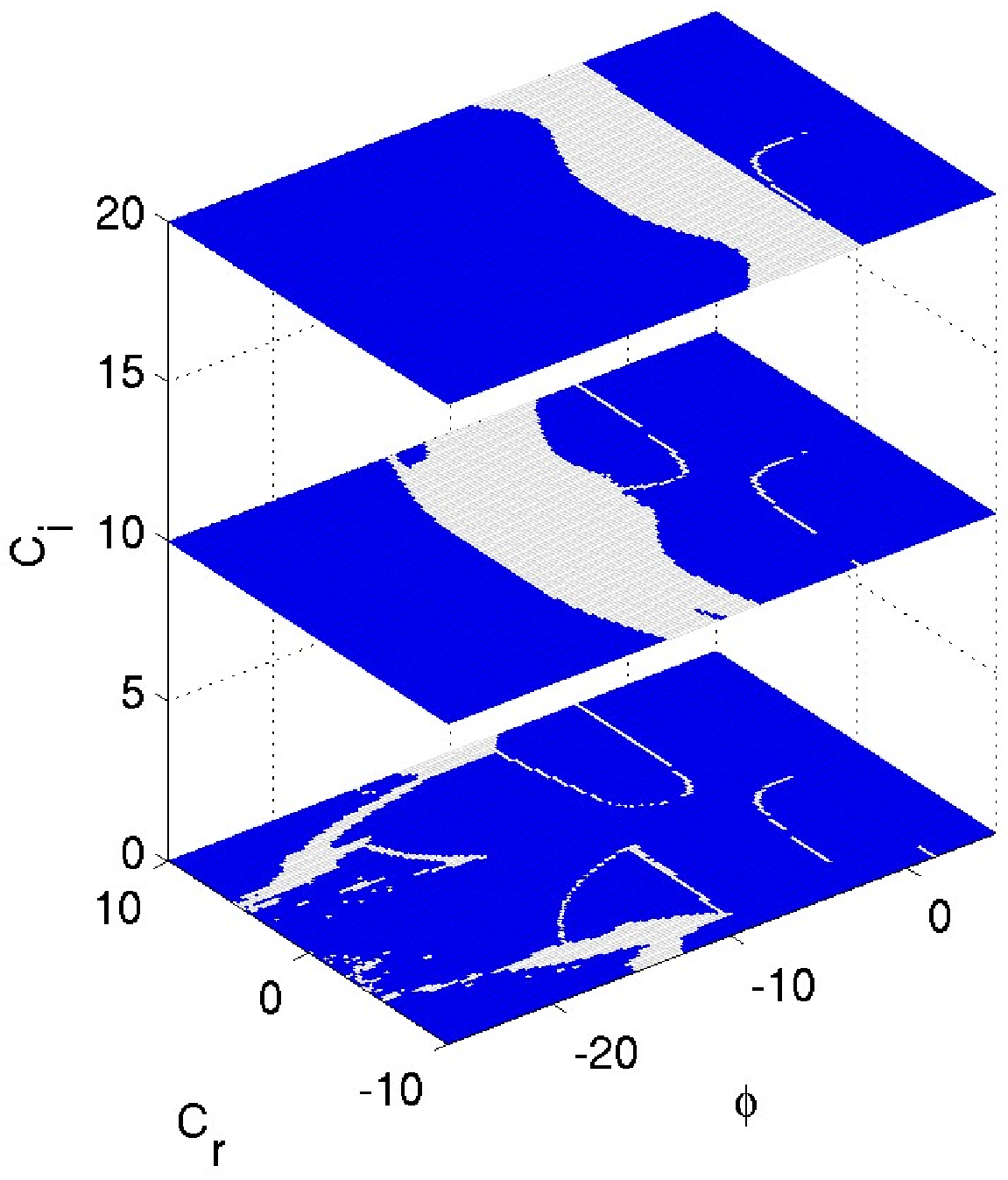} 
\\
    \includegraphics[width=0.5\textwidth]{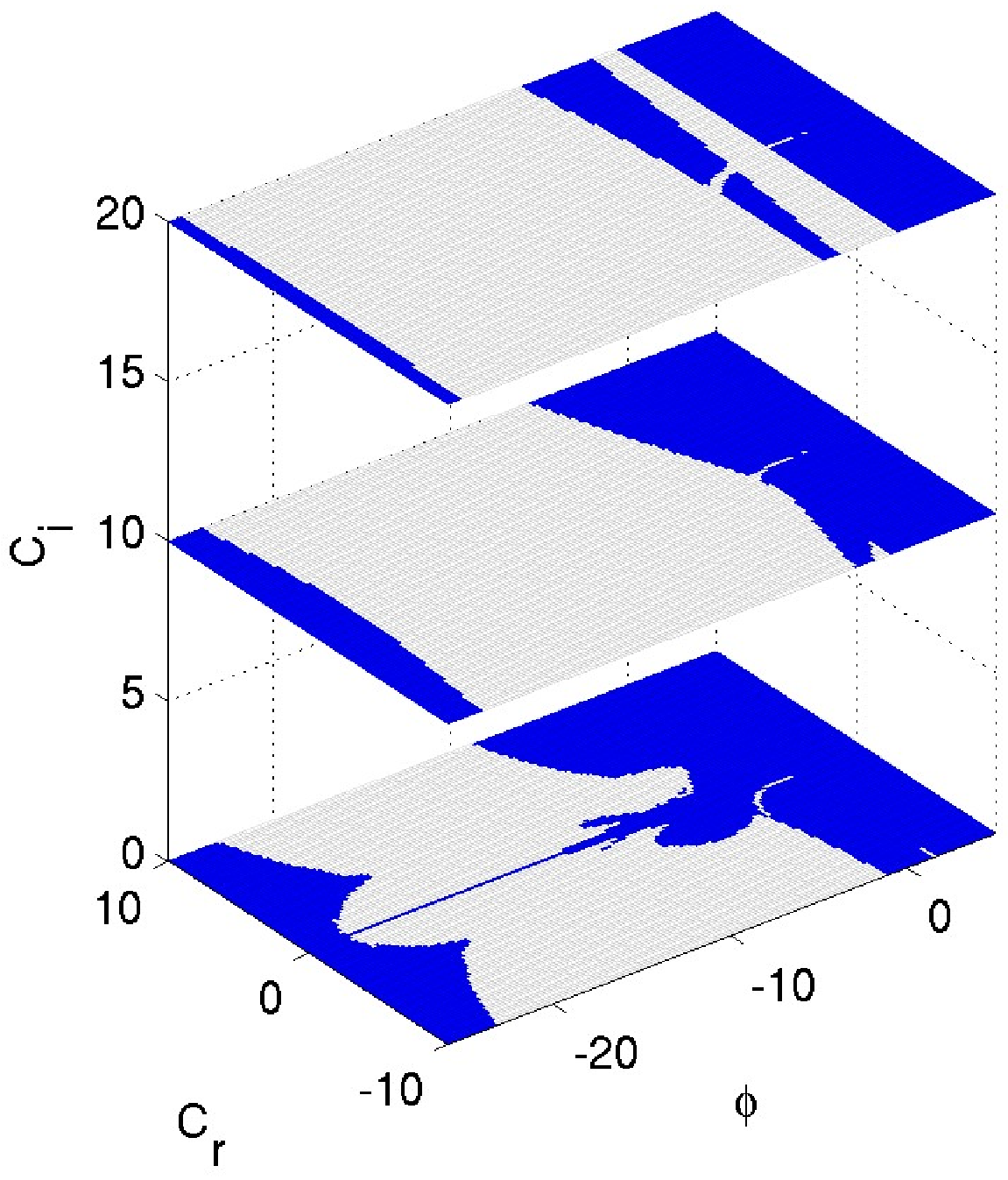}
&
    \includegraphics[width=0.5\textwidth]{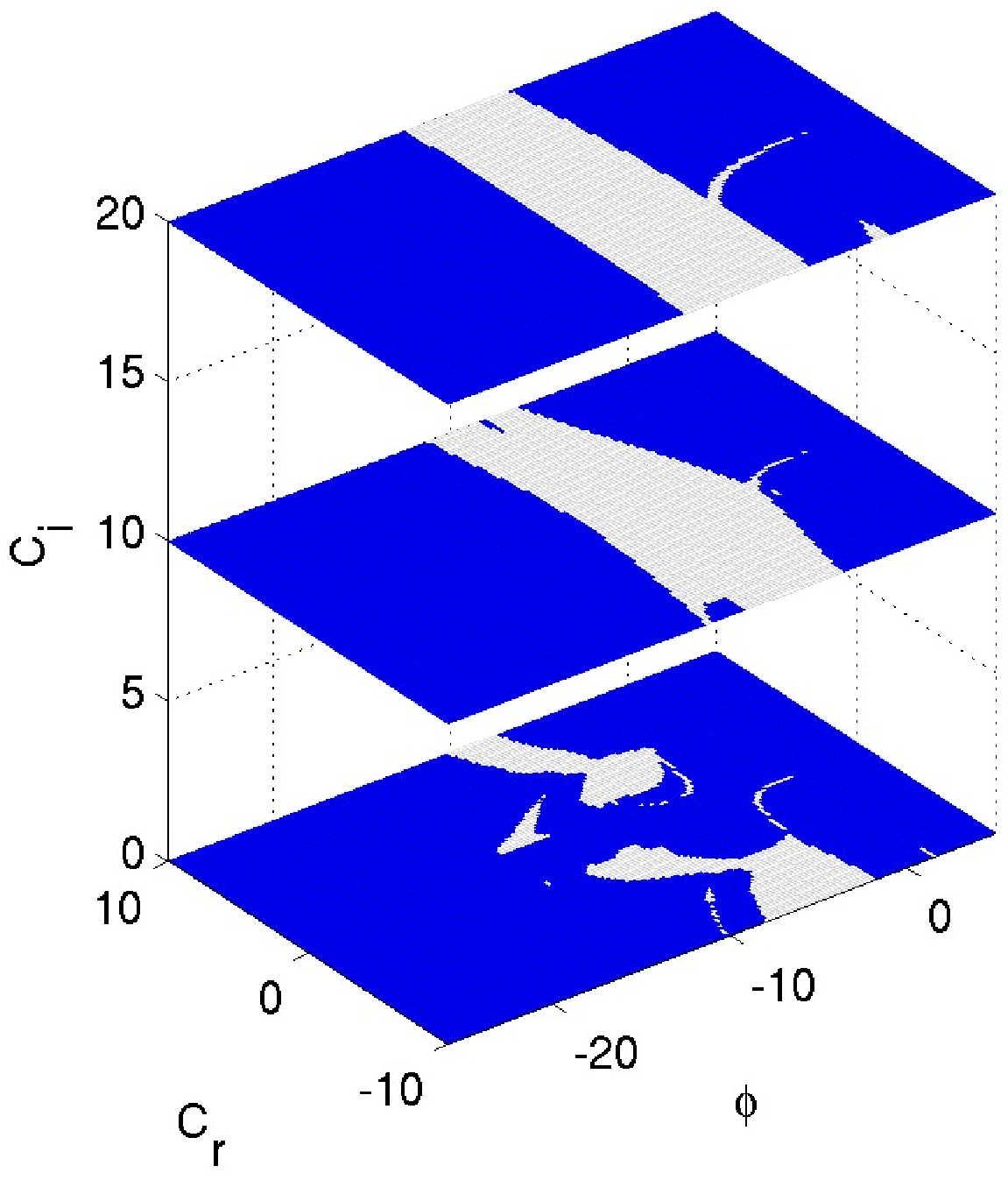} 
\\
    \includegraphics[width=0.5\textwidth]{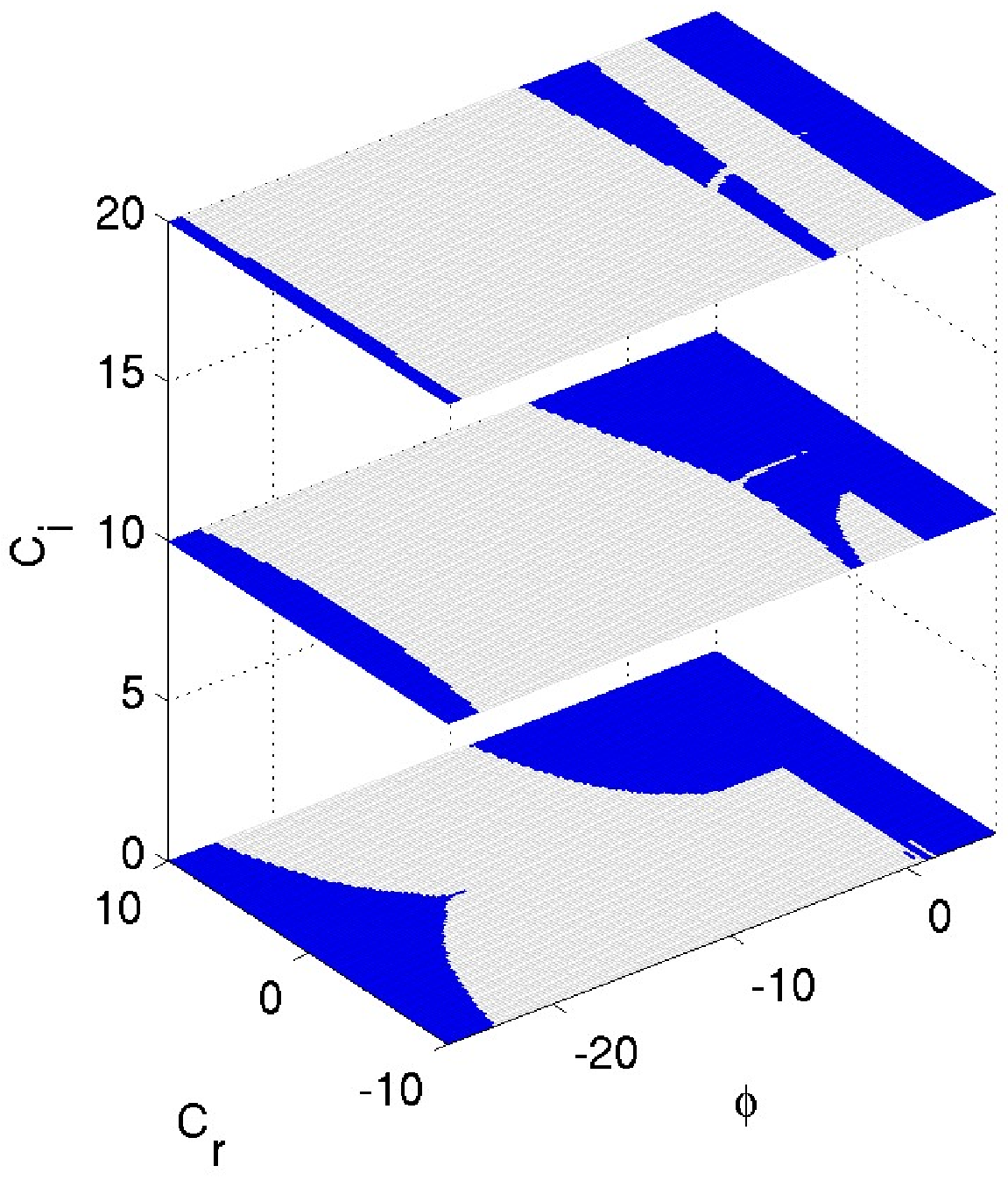}
&
    \includegraphics[width=0.5\textwidth]{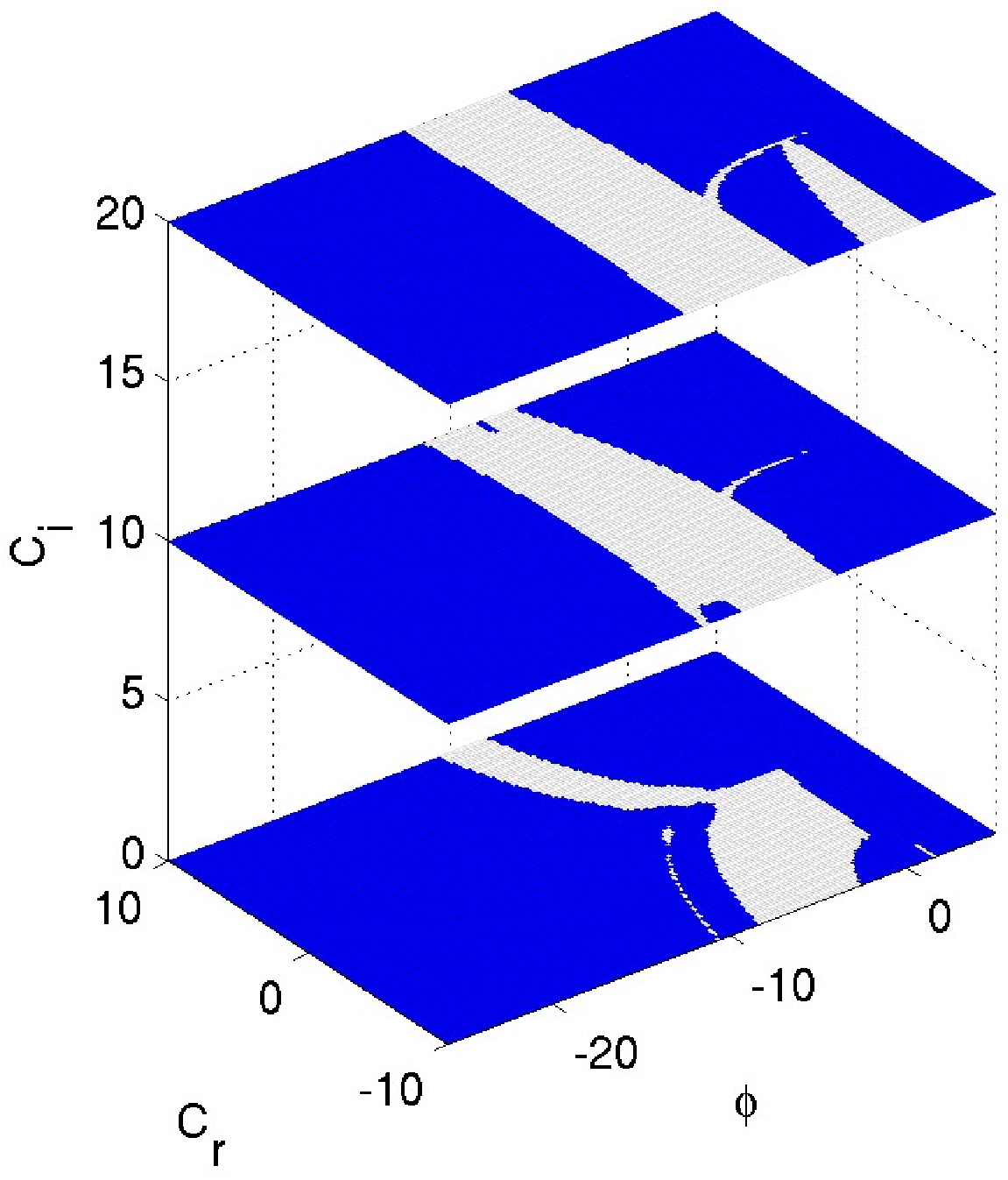} 
\\
  \end{tabular}
  \end{center}
\caption{Matter (left column) and  Radiation (right column), from top to bottom: $\Omega_{b0}=10^{-10},10^{-2},0.9$.  A measure of stabilisation is the percentage of the total area which leads to stabilisation.  Values are given in Table \ref{tab-areas}.}
\label{fig-vol}
\end{figure}

It is convenient to define a measure of goodness of stability.  Firstly, to determine whether the background fluid adds stabilisation, we define an overall normalised, 4-D volume, ${\cal V}$, which is given by the volume of stabilised trajectories as a percentage of the total volume.  In addition, we may calculate the ratio of stabilised trajectories to the total trajectories on a given 2-D plane (again, as a percentage), ${\cal A}_{ab}$, where $a$ and $b$ denote the fixed values of the remaining fields.  (For example ${\cal A}_{C_{i}=10,T_{i}=0}$ represents the stabilisation area of the $\phi$-$C_{r}$ plane when $C_{i0}=10$ and $T_{i0}=0$).

The numerical results for the measure of goodness are given in Tables \ref{tab-areas} and \ref{tab-vols}.
It should be well noted that the numbers quoted are intended to show a trend and are only indicative.  
Numerically, we rely on a finite volume, with finite (coarse) spacing. We chose a grid size that was numerically feasible within timescales.  The may preclude us from resolving all the fine structure.  
We do note, however, that the fine structure arises from the asymmetry of the potential and the fine-tuning of the trajectories.
This will be more relevant when we consider a radiation background fluid.
\begin{table}[t!]
  \begin{center}
   \begin{tabular}{|cc|cc|}
   \hline 
  &  &Matter & Radiation \\
   $\Omega_{b0}$ & $C_{i}$ & ${\cal A}_{C_{i},T_{i}=0}$ (\%) & ${\cal A}_{C_{i},T_{i}=0}$ (\%) \\
   \hline
   $10^{-10}$ & $20$ & $68$ & $18$\\
    & $10$ & $65$ & $24$\\
    & $0$ & $39$ & $9$\\ \hline
   $10^{-2}$ & $20$ & $71$ & $22$ \\
    & $10$ & $66$ & $22$\\
    & $0$ & $54$ & $11$\\ \hline
   $0.9$ & $20$ & $77$ & $25$\\
    & $10$ & $67$ & $21$\\
    & $0$ & $59$ & $15$\\ \hline
   \end{tabular}
   \end{center}
\caption{Table showing the stabilisation measure, ${\cal A}_{ab}$, (percentage of the total area which leads to stabilisation) as shown graphically in Figure \ref{fig-vol}. These numbers are indicative of trends only, as explained in the text.}
\label{tab-areas}
\end{table}
\begin{table}[t!]
  \begin{center}
   \begin{tabular}{|c|cc|}
   \hline
   &Matter & Radiation \\
   $\Omega_{b0}$ &  ${\cal V}$ (\%) & ${\cal V}$ (\%) \\
   \hline
   $10^{-10}$ & $42$ & $13$\\ \hline
   $10^{-2}$ & $48$ & $15$ \\ \hline
   $0.9$ &  $52$ & $18$\\ \hline
   \end{tabular}
   \end{center}
\caption{Table showing the stabilisation measure, ${\cal V}$, (percentage of the total 4-D volume which leads to stabilisation).  2-D cross-sections are shown graphically in Figure \ref{fig-vol}. These numbers are indicative of trends only, as explained in the text.}
\label{tab-vols}
\end{table}

Our results indicate that a background matter fluid aids stabilisation, consistent with previous work~\cite{Barreiro1,Brustein:2004jp,Barreiro:2005ua}.  In addition, the presence of matter fields further aids stabilisation.  Importantly, while a non-zero value of $C_r$ assists the stabilisation of $\phi$, the imaginary part $C_i$ further increases the stabilsation regions.

\section{Radiation ($\gamma=\frac43$)}
\label{secRad}
For completeness, we briefly consider the effects of a radiation background fluid, in place of a matter fluid.  
All the equations in the preceding sections are valid, where $\gamma=4/3$.
The results are shown in the righthand column of Fig.~\ref{fig-vol} and Tables~\ref{tab-areas} and~\ref{tab-vols}.
The main conclusions are:
\begin{enumerate}
\item For all scenarios, the stabilisation region is much reduced, compared to when $\gamma=1$.  
\item Increasing $\Omega_{b0}$ again aids stabilisation.
\item Due to the reduced cosmological friction acting on the fields, more fine-tuning is necessary for trajectories to find the local minimum.
\item We observe the same trend of stabilisation when $C_r$ and $C_i$ are non-zero as for the matter background fluid, although the effects are less pronounced.
\end{enumerate}

\section{Conclusion}
\label{secConc}
In this paper, we have considered moduli stabilisation in the KKLT scenario coupled to a Polonyi matter field, a model which has been previously studied in the context of particle phenomenology.  
In this scenario, the AdS vacuum is uplifted by F-terms in supergravity.
We have examined the evolution of two complex fields (one modulus and one matter field), taking also a background fluid into account, either matter or radiation.
We find that the presence of both the background fluid and the matter field enlarge the region leading to stabilisation of the moduli fields, due to a new scaling regime. 
Although a more detailed treatment is necessary, we believe that our conclusions will hold for similar scenarios, where the new direction is steep.
A matter background fluid aids stabilisation more than radiation, consistent with previous studies. 

\acknowledgments
The authors acknowlege the use of the package ``SuperCosmology'' to calculate the scalar potential~\cite{Kallosh:2004rs} and would like to thank STFC for financial support.


\vspace{5mm}
\appendix
\noindent {\bf\Large Appendix}

\section{Useful Formulae}
\label{App2}
From the definition of $\lambda$, $\delta$, $\eta$ and $\theta$, one obtains
\begin{align*}
\frac{\lambda^\prime}{\lambda}&=\sqrt{6} x \lambda \big[1-\Gamma_{\phi\phi}\big] 
	+ \sqrt{6} z \eta \big[1-\Gamma_{\phi T_i}\big]
	+ \sqrt{3} p \delta \big[1-\Gamma_{\phi C_r}\big]
	+ \sqrt{3} q \theta \big[1-\Gamma_{\phi C_i}\big], \\
\frac{\delta^\prime}{\delta}&=\sqrt{6} x \lambda \big[1-\Gamma_{\phi~C_r}\big] 
	+ \sqrt{6} z \eta \big[1-\Gamma_{C_r T_i}\big]
	+ \sqrt{3} p \delta \big[1-\Gamma_{C_r C_r}\big]
	+ \sqrt{3} q \theta \big[1-\Gamma_{C_r C_i}\big], \\
\frac{\eta^\prime}{\eta}&=2x+ 
          \sqrt{6} x \lambda \big[1-\Gamma_{\phi T_i}\big] 
	+ \sqrt{6} z \eta \big[1-\Gamma_{T_i T_i}\big]
	+ \sqrt{3} p \delta \big[1-\Gamma_{T_i C_r}\big]
	+ \sqrt{3} q \theta \big[1-\Gamma_{T_i C_i}\big], \\
\frac{\theta^\prime}{\theta}&=\sqrt{6} x \lambda \big[1-\Gamma_{\phi C_i}\big] 
	+ \sqrt{6} z \eta \big[1-\Gamma_{C_i T_i}\big]
	+ \sqrt{3} p \delta \big[1-\Gamma_{C_r C_i}\big]
	+ \sqrt{3} q \theta \big[1-\Gamma_{C_i C_i}\big],
\end{align*}
where we define
\begin{eqnarray*}
\Gamma_{ij} \equiv \frac{V V_{ij}}{V_i V_j}.
\end{eqnarray*}
We take a change of variables
\begin{eqnarray*}
\epsilon_\phi\equiv\frac{1}{\lambda} , \quad \epsilon_{T_i}\equiv\frac{1}{\eta} , \quad  \epsilon_{C_r} 
\equiv\frac{1}{\delta}, \quad \epsilon_{C_i}\equiv\frac{1}{\theta} ,
\end{eqnarray*}
and we define $X$, $Y$, $Z$, $P$ and $Q$ to be
\begin{eqnarray*}
x = \epsilon_\phi X , \quad y = \epsilon_{\phi} Y,\quad z = \epsilon_{T_i} Z, \quad p = 
\epsilon_{C_r} P, \quad q = \epsilon_{C_i} Q.
\end{eqnarray*}

With these new variables we obtain:
\begin{align*}
\begin{split}
\frac{H^\prime}{H} = -\frac32 \big[2\epsilon_\phi^2X^2 + 2\epsilon_{T_i}^2 &Z^2 + 2\epsilon_{C_r}^2P^2+ 2\epsilon_{C_i}^2Q^2 \\
&+ \gamma\left(1-\epsilon_\phi^2X^2-\epsilon_\phi^2Y^2 -\epsilon_{T_i}^2 Z^2 -
\epsilon_{C_r}^2P^2-\epsilon_{C_i}^2Q^2 \right) \big]
\end{split}
\end{align*}
and
\begin{eqnarray*}
	X^{\prime} &=& \frac{\lambda^\prime}{\lambda} X - 3X +\sqrt{\frac32} Y^2 - 2 \frac{\epsilon_{T_i}^2}
{\epsilon_\phi} Z^2 - \frac{H^{\prime}}{H} X \\
Y^{\prime} &=& \frac{\lambda^\prime}{\lambda} Y  -\sqrt{\frac32} XY -\sqrt{\frac32} ZY -\frac{\sqrt{3}}{2} PY 
-\frac{\sqrt{3}}{2} QY  - \frac{H^{\prime}}{H} Y \\
Z^{\prime} &=& \frac{\eta^\prime}{\eta} Z - 3Z + \sqrt{\frac32}\left(\frac{\epsilon_{\phi}}{\epsilon_{T_i}}\right)
^2 Y^2 + 2 \epsilon_\phi XZ - \frac{H^{\prime}}{H} Z \\
P^{\prime} &=& \frac{\delta^\prime}{\delta} P - 3P + \frac{\sqrt{3}}{2}\left(\frac{\epsilon_{\phi}}{\epsilon_
{C_r}}\right)^2 Y^2 - \frac{H^{\prime}}{H} P \\
Q^{\prime} &=& \frac{\theta^\prime}{\theta} Q - 3Q + \frac{\sqrt{3}}{2}\left(\frac{\epsilon_{\phi}}{\epsilon_
{C_i}}\right)^2 Y^2 - \frac{H^{\prime}}{H} Q 
\end{eqnarray*}
Finally, from above, we find
\begin{align}
\frac{\epsilon_\phi^\prime}{\epsilon_\phi}&=-\left\{\sqrt{6} X \big[1-\Gamma_{\phi\phi}\big] 
	+ \sqrt{6} Z \big[1-\Gamma_{\phi T_i}\big]
	+ \sqrt{3} P \big[1-\Gamma_{\phi C_r}\big]
	+ \sqrt{3} Q \big[1-\Gamma_{\phi C_i}\big]\right\} \label{eps_phi_prime}\\
\frac{\epsilon_{C_r}^\prime}{\epsilon_{C_r}}&=-\left\{\sqrt{6} X \big[1-\Gamma_{\phi C_r}\big] 
	+ \sqrt{6} Z \big[1-\Gamma_{C_r T_i}\big]
	+ \sqrt{3} P \big[1-\Gamma_{C_r C_r}\big]
	+ \sqrt{3} Q \big[1-\Gamma_{C_r C_i}\big]\right\} \label{eps_Cr_prime}\\
\frac{\epsilon_{T_i}^\prime}{\epsilon_{T_i}}&=-\left\{2\epsilon_\phi X+
          \sqrt{6} X \big[1-\Gamma_{\phi T_i}\big] 
	+ \sqrt{6} Z \big[1-\Gamma_{T_i T_i}\big]
	+ \sqrt{3} P \big[1-\Gamma_{T_i C_r}\big]
	+ \sqrt{3} Q \big[1-\Gamma_{T_i C_i}\big]\right\} \label{eps_Ti_prime}\\
\epsilon_{C_i}^\prime&=-\epsilon_{C_i}\left\{\sqrt{6} X \big[1-\Gamma_{\phi C_i}\big] 
	+ \sqrt{6} Z \big[1-\Gamma_{C_i T_i}\big]
	+ \sqrt{3} P \big[1-\Gamma_{C_r C_i}\big]
	+ \sqrt{3} Q \big[1-\Gamma_{C_i C_i}\big] \right\} \label{eps_Ci_prime}
\end{align}

\section{Tracking Solution}
\label{App1}
In the following, we generalise the scaling regime found in \cite{Ng:2001hs} for two real fields.  We 
therefore consider $T_i=C_i=0$ in the following.
The simplified equations of motion considered (with $Q=Z=0$) are 
\begin{eqnarray}
X^{\prime} =& -\sqrt6(\Gamma_{\phi\phi}-1)X^2& -\sqrt3(\Gamma_{\phi C_r}-1)XP - 3X 
      +\sqrt{\frac32} Y^2 
\label{EOMX}\\
     && +  \frac32X\left[2(\epsilon_\phi^2 X^2+\epsilon_{C_r}^2P^2) + \gamma(1- \epsilon_\phi^2 X^2-
\epsilon_\phi^2 Y^2-\epsilon_{C_r}^2P^2 ) \right], \nonumber \\
Y^{\prime} =& -\sqrt6(\Gamma_{\phi\phi}-1)XY & -\sqrt3(\Gamma_{\phi C_r}-1)YP
  -\sqrt{\frac32} XY  -\frac{\sqrt{3}}{2} PY  
\label{EOMY} \\
     && +\frac32Y\left[2(\epsilon_\phi^2 X^2+\epsilon_{C_r}^2P^2) + \gamma(1- \epsilon_\phi^2 X^2-
\epsilon_\phi^2 Y^2-\epsilon_{C_r}^2P^2 ) \right], \nonumber  \\
P^{\prime} =&  -\sqrt6(\Gamma_{\phi C_r}-1)XP &-\sqrt3(\Gamma_{C_r C_r}-1)P^2 - 3P +\frac{\sqrt{3}}{2}
\frac{\epsilon_\phi^2}{\epsilon_{C_r}^2} Y^2  
\label{EOMP} \\
     && +  \frac32P\left[2(\epsilon_\phi^2 X^2+\epsilon_{C_r}^2P^2) + \gamma(1- \epsilon_\phi^2 X^2-
\epsilon_\phi^2 Y^2-\epsilon_{C_r}^2P^2 ) \right]. \nonumber
\end{eqnarray}
Small $\epsilon_\phi$ and $\epsilon_
{C_r}$ or  $\Gamma_{ij}\approx1$ imply that $\epsilon_i$ is almost constant, as seen in Equations (\ref{eps_phi_prime}) and (\ref{eps_Cr_prime}).
Therefore the ``instant critical point'' is obtained from solving $X^{\prime}=Y^{\prime}=P^{\prime}=0$.
From Eqns.~(\ref{EOMX})-(\ref{EOMP}), we obtain
\begin{eqnarray}
&Y^2=-X^2+\sqrt6 X-\frac{1}{\sqrt2}XP, \label{App_Y^2}\\
&Y^2(P-\frac{\epsilon_\phi^2}{\sqrt2 \epsilon_{C_r}^2}X)
=2(\Gamma_{\phi\phi}-\Gamma_{\phi C_r})X^2P +\sqrt2(\Gamma_{\phi C_r}-\Gamma_{C_r C_r})XP^2.
\end{eqnarray}
It is reasonable to assume (in the regions we consider in the paper)
\begin{eqnarray*}
\Gamma_{\phi\phi}\simeq\Gamma_{C_r C_r}\simeq \Gamma_{\phi C_r}\simeq 1,
\end{eqnarray*}
from which we find 
\begin{eqnarray*}
P\simeq\frac{\epsilon_\phi^2}{\sqrt2\epsilon_{C_r}^2}X.
\end{eqnarray*}
Plugging this solution into Eqn.~(\ref{App_Y^2}), both $P=P(X)$ and $Y=Y(X)$.  Solving Eqn.~(\ref
{EOMX}) (using $X^{\prime}=0$) leads to
\dis{
x_c=\sqrt{\frac{3}{2}}
\frac{\tilde\gamma}{\lambda},\qquad p_c=\frac{\sqrt3}{2}\frac{\delta\tilde\gamma}{\lambda^2},
\qquad y_c^2=\frac32\frac{\tilde\gamma}{\lambda^2}\left(2-\tilde\gamma-\frac{\delta^2}{2\lambda^2}
\tilde\gamma \right),
\label{scalapp}
}
where 
\dis{
\tilde\gamma=\gamma\left(1+\frac{\delta^2}{2\lambda^2} \right)^{-1}.
\label{gamtilde}
}

\section{Stability of the critical points}
\label{AppStability}
We expand about the critical points
\disn{
X=X_c+u,\qquad Y=Y_c+v,\qquad P=P_c+w,
}
which yield, to first order, the equations of motion
\disn{
\left(\begin{array}{c} u^\prime \\ v^\prime \\w^\prime \end{array}\right)
=M \left(\begin{array}{c} u \\ v\\w \end{array}\right).
}
Assuming $\Gamma$'s are 1 (this assumption is quite good
for the region of approximately $\phi<-8$ and $|C_r|>1$), 
we find the eigenvalues, $m_i$, of $M$.
We also assume $0\leq\gamma\leq 2$ for the baryotropic fluid.

\noindent{\underline{\it Fluid-dominated solution}} \\
For this solution, $x_c=p_c=y_c=0$. We find a saddle point for $0<\gamma<2$:
\disn{
m_1=\frac{3}{2}\gamma,\qquad m_2=m_3=-\frac{3}{2}(2-\gamma).
}

\vspace{1mm}
\noindent{\underline{\it Kinetic-dominated solutions}} \\
Here, $x_c^2+p_c^2=1$ and $y_c=0$.
We find an unstable node for $\lambda x_c+(\delta/\sqrt2)p_c < \sqrt6$ and a saddle point for $\lambda x_c+(\delta/\sqrt2)p_c> \sqrt6$:
\disn{
m_1=0,\qquad m_2=3(2-\gamma),\qquad m_3=\sqrt{\frac32}\left(\sqrt6-\lambda x_c-\frac{\delta p_c}{\sqrt2}\right).
}

\vspace{1mm}
\noindent{\underline{\it Scalar field dominated solution}} \\
In this case, $x_c=\frac{\lambda}{\sqrt6}$, $p_c=\frac{\delta}{2\sqrt3}$, 
$y_c^2=1-\frac{1}{12}\left(2\lambda^2+\delta^2\right)$.
 There is a stable node for $\lambda^2+\delta^2/2<3\gamma$ and a saddle point for $3\gamma<\lambda^2+
\delta^2/2<6$:
\disn{
m_{1,2}=-3+\frac14(\delta^2+2\lambda^2),\qquad 
m_3=-3\gamma+\frac12(\delta^2+2\lambda^2).
}

\vspace{1mm}
\noindent{\underline{\it Scaling solution}} \\
For the scaling solution, $x_c$, $y_c$ and $p_c$ are given in Eqns.~(\ref{scalapp}) and~(\ref{gamtilde}).
 We find a stable node for $3\gamma<\lambda^2+\delta^2/2<24\gamma^2/(9\gamma-2)$ and a stable spiral for $\lambda^2>24\gamma^2/(9\gamma-2)$:

\disn{
m_1&=-\frac{3}{2}(2-\gamma),\\
m_{2,3}&=-\frac34(2-\gamma)\Big[1\pm\sqrt{1-\frac{8\gamma(\lambda^2+\delta^2/2-3\gamma))}
{(\lambda^2+\delta^2/2)(2-\gamma)}}\Big].
}

\end{document}